\documentclass[twocolumn,showpacs,aps,prd,superscriptaddress]{revtex4} 

\usepackage{graphicx}
\usepackage{dcolumn}
\usepackage{amsmath}
\usepackage{amssymb}
\usepackage{latexsym,mathrsfs}
\usepackage{epsfig}
\usepackage{bm}
\newcommand{\Rab}{R_{12}}
\newcommand{\Rbc}{R_{23}}
\newcommand{\Rac}{R_{13}}
\newcommand{\Wab}{W_{12}}
\newcommand{\Wbc}{W_{23}}
\newcommand{\Wac}{W_{13}}

\newcommand{\nue}{\nu_{e}}
\newcommand{\numu}{\nu_{\mu}}
\newcommand{\nutau}{\nu_{\tau}}
\newcommand{\Dmab}{\Delta m^{2}_{21}}

\newcommand{\Dmbc}{\Delta m^{2}_{32}}
\newcommand{\sinthbc}{\sin^{2}\theta_{23}}
\newcommand{\tanthab}{\tan^{2}\theta_{12}}
\newcommand{\sinthac}{\sin^{2}\theta_{13}}

\newcommand{\RepA}{\mathcal{A}}
\newcommand{\RepD}{\mathcal{D}}
\newcommand{\RepH}{\mathcal{H}}
\newcommand{\RepG}{\mathcal{G}}
\newcommand{\CPV}{\delta_{cp}}
\newcommand{\CPVA}{\delta_{cp}^{\mathcal{A}} }
\newcommand{\CPVD}{\delta_{cp}^{\mathcal{D}} }
\newcommand{\CPVH}{\delta_{cp}^{\mathcal{H}} }
\newcommand{\cA}{c^{\mathcal{A}} }
\newcommand{\cD}{c^{\mathcal{D}} }
\newcommand{\sA}{s^{\mathcal{A}} }
\newcommand{\sD}{s^{\mathcal{D}} }
\newcommand{\PhiA}{\Phi^{\mathcal{A}} }

\newcommand{\rhoD}{\rho^{\mathcal{D}} }
\newcommand{\rhoH}{\rho^{\mathcal{H}} }

\newcommand{\thH}{\theta^{\mathcal{H}} }
\newcommand{\thG}{\theta^{\mathcal{G}} }
\newcommand{\thA}{\theta^{\mathcal{A}} }
\newcommand{\thD}{\theta^{\mathcal{D}} }
\newcommand{\thDA}{\theta^{\mathcal{DA}} }
\newcommand{\DAmij}{\Delta^{\mathcal{A}} m^{2}_{ij}}
\newcommand{\DDmij}{\Delta^{\mathcal{D}} m^{2}_{ij}}

\newcommand{\DAmab}{\Delta^{\mathcal{A}} m^{2}_{21}}

\newcommand{\DAmbc}{\Delta^{\mathcal{A}} m^{2}_{32}}
\newcommand{\DDmab}{\Delta^{\mathcal{D}} m^{2}_{21}}

\newcommand{\DDmbc}{\Delta^{\mathcal{D}} m^{2}_{32}}

\newcommand{\sinthbcA}{\sin^{2}\thA_{23}}
\newcommand{\tanthabA}{\tan^{2}\thA_{12}}
\newcommand{\sinthacA}{\sin^{2}\thA_{13}}

\newcommand{\chitwo}{\chi^{2}}
\newcommand{\barnue}{\bar{\nu}_e}  

\begin{document}


\title{
\begin{center}
{\Large \textbf{
     Global Neutrino Data Analysis and the Quest to Pin \vspace{0.2cm} Down \\ 
     $\sin\theta_{13}$ in Different Mixing Matrix Parametrizations }}
\end{center}
}

\affiliation{Leung Center for Cosmology and Particle Astrophysics,
             National Taiwan University, Taipei 10617, Taiwan}
\affiliation{Department of Physics, Laurentian University, Sudbuary, 
             ON P3E 2C6, Canada}
\affiliation{Fermi National Accelerator Laboratory, Batavia, IL 60510, USA}
\affiliation{Department of Physics, University of Guelph, Guelph, 
             ON N1G 2W1, Canada}
\affiliation{Graduate Institute of Physics, National Taiwan University,
             Taipei 10617, Taiwan}

\author{Melin Huang}
   \affiliation{Leung Center for Cosmology and Particle Astrophysics,
                National Taiwan University, Taipei 10617, Taiwan}
   \affiliation{Department of Physics, Laurentian University, Sudbuary, 
                ON P3E 2C6, Canada}
\author{S. D. Reitzner}
   \affiliation{Fermi National Accelerator Laboratory, Batavia, IL 60510, USA}
   \affiliation{Department of Physics, University of Guelph, 
                Guelph, ON N1G 2W1, Canada}
\author{Wei-Chun Tsai}
   \affiliation{Graduate Institute of Physics, National Taiwan University,
             Taipei 10617, Taiwan} 
\author{Huitzu Tu}
   \affiliation{Leung Center for Cosmology and Particle Astrophysics,
                National Taiwan University, Taipei 10617, Taiwan}

\date{\today}

\begin{abstract}
\vspace{0.1in}
Hints for sizable $\sin^2 \theta_{13}$ have been reported in earlier global 
neutrino oscillation data analyses as well as will be reported in this work, 
and quite recently by the Double Chooz experiment.  However, as we enter the 
era of precision neutrino oscillation experiments, terms linear in 
$\sin\theta_{13}$ will no longer be negligible, and its sign would affect 
the extraction of other oscillation parameters.  The sign of $\sin\theta_{13}$ 
also plays a crucial role in the determination of the CP-violating phase.
In this work we show that by adopting an alternative parametrization for 
the Pontecorvo-Maki-Nakagawa-Sakata (PMNS) mixing matrix, one already has 
a chance to infer the sign of each mixing angle in the conventional 
parametrization using existing global neutrino data. 
A weak preference for negative $\sin \theta_{13}$ is found. 
In particular, the solar data suggest that $\sin\theta_{13} > 0$ while 
all other data the opposite. This leads to the speculation on whether 
the Mikheyev-Smirnov-Wolfenstein (MSW) effect is responsible.
In this work we found that in the new mixing matrix parametrization, 
the $68\%$ CL constraints on the three mixing angles are comparable 
to those estimated in the conventional parametrization adopted in the 
literature. Owing to the strong correlations among the three mixing angles
in the new parametrization, the advantages of doing the global neutrino 
oscillation analysis using data from past, current, and near future 
neutrino oscillation experiments shall become manifest.
\end{abstract}


\maketitle


%
\section{Introduction}\label{sec:intro}

The experiments involving solar, reactor, atmospheric and accelerator 
neutrinos have established a picture of neutrino oscillations caused 
by non-zero neutrino masses and mixing among different neutrino flavors 
(see e.g. Ref.~\cite{Nakamura:2010zzi}). 
The phenomenology of neutrino oscillations arising from the mismatch 
between the weak and the mass eigenstates can be described by the 
Pontecorvo-Maki-Nakagawa-Sakata (PMNS) mixing matrix~\cite{Maki:1962mu, 
Pontecorvo:1967fh}. 
This matrix can be parametrized in various ways as seen in the 
literature (e.g., \cite{Schechter:1980gr,Fritzsch:1997st,Choubey:2000bf,
Giunti_Kim,Zheng:2010kp,Huang:2011by}).

All experimental data except those from the LSND~\cite{Aguilar:2001ty} can be
well described assuming three active neutrinos. In the case of Dirac neutrinos,
the $3 \times 3$ unitary mixing matrix is characterized by three Euler angles 
and one physical phase, and can be expressed as a product of three rotation 
matrices. The physical phase can be responsible for CP violation in the 
neutrino sector. As mentioned in Ref.~\cite{Fritzsch:1997st}, the CP-violating 
phase can be associated with the sine or cosine of any mixing angle or with 
the identity entry in any of the three rotation matrices. In this work, 
we will follow the standard Cabibbo-Kobayashi-Maskawa (CKM) 
matrix~\cite{Chau:1984fp} for the assignment of the CP-violating phase.  

Define
\begin{equation*}
R_{23} = 
  \left(
     \begin{array}{ccc}
       1         & 0          & 0       \\
       0         &  c_{23}    & s_{23}  \\
       0         & -s_{23}    & c_{23} 
     \end{array}
  \right),  \hskip0.5cm
R_{13} = 
  \left(
     \begin{array}{ccc}
       c_{13}    & 0          & s_{13}   \\
       0         & 1          & 0        \\
      -s_{13}    & 0          & c_{13} 
     \end{array}
  \right),  \vspace{-0.3cm} 
\end{equation*}
\begin{equation}
R_{12} = 
  \left(
     \begin{array}{ccc}
       c_{12}    & s_{12}     & 0        \\
      -s_{12}    & c_{12}     & 0        \\
       0         & 0          & 1 
     \end{array}
  \right),   \vspace{-0.3cm}   
\label{eq:Gfit_01}
\end{equation}
and
\begin{equation*}
W_{23} =  
  \left(
    \begin{array}{ccc}
       1   &   0                  &   0                  \\
       0   &   c_{23}             &   s_{23} e^{-i\CPV}  \\
       0   &  -s_{23} e^{i\CPV}   &   c_{23} 
    \end{array}    
  \right),        \nonumber \\ 
\end{equation*}
\begin{equation*}
W_{13} =  
  \left(
    \begin{array}{ccc}
       c_{13}             &   0   &   s_{13} e^{-i\CPV}  \\
       0                  &   1   &   0                  \\
      -s_{13} e^{i\CPV}   &   0   &   c_{13} 
    \end{array}
  \right),         \nonumber \\ 
\end{equation*}
\begin{equation}
W_{12} =  
  \left(
    \begin{array}{ccc}
       c_{12}             &   s_{12} e^{-i\CPV}   &  0   \\
      -s_{12} e^{i\CPV}   &   c_{12}              &   0  \\
       0                  &   0                   &   1 
    \end{array}
  \right), 
\label{eq:Gfit_02}
\end{equation}
where $\theta_{ij}$ and $\CPV$ are the mixing angles and the CP-violating 
phase, respectively, $c_{ij} \equiv \cos\theta_{ij}$ and 
$s_{ij} \equiv \sin\theta_{ij}$.
As the combination $R_{23}W_{13}R_{12}$ is the standard choice for describing 
the quark mixing, it has been adopted to be the conventional parametrization
for the mixing matrix for Dirac neutrinos. This choice of the parametrization 
was actually made prior to the era when neutrino oscillation data became 
available.  Later, the solar neutrino 
experiments~\cite{Aharmim:2009gd,Aharmim:2011vm,Hosaka:2005um,Cravens:2008zn}, 
reactor~\cite{Gando:2010aa,Apollonio:2002gd}, long-baseline (LBL) 
accelerator~\cite{Ahn:2006zza,Adamson:2011ig,MINOS_nue_app_2011}, 
as well as the atmospheric neutrino 
experiments~\cite{Wendell:2010md,Aharmim:2009zm,Hosaka:2006zd}
all choose this parametrization to present their results.
Using existing global neutrino data, most phenomenology 
works~\cite{Fogli:2011qn,Schwetz:2011qt,GonzalezGarcia:2010er,Roa:2009wp, 
Balantekin:2008zm,Ge:2008sj,Goswami:2004cn,Choubey:2003uw} also employ this 
parametrization to determine neutrino oscillation parameters.

In the conventional parametrization, the three mixing angles happen to nearly 
decouple for solar, atmospheric/accelerator, and reactor neutrino oscillation 
experiments due to the different neutrino energies and traveling distances 
involved.  The $\theta_{12}$ and $\theta_{23}$ angles are well determined 
by solar and atmospheric experiments, respectively.  Current-generation 
short-baseline reactor experiments such as the Daya Bay~\cite{Guo:2007ug}, 
Double Chooz~\cite{Ardellier:2006mn}, RENO~\cite{Joo:2007zzb} and the 
Angra~\cite{Anjos:2005pg} experiments are exploited to pin down the yet 
unknown $\theta_{13}$ value.
Non-zero or sizable $\theta_{13}$ values are predicted by many neutrino mass 
models (see e.g. Ref.~\cite{Chen:2008eq} for a nice compilation of existing 
model predictions), and supported by global neutrino data analyses (see e.g.
Ref~\cite{Fogli:2011qn,Schwetz:2011qt,GonzalezGarcia:2010er,Ge:2008sj} and 
this work).  
Very recently, the Double Chooz experiment~\cite{double_chooz_2011} has 
reported their preliminary results of 
$\sin^2 2\theta_{13} = 0.085 \pm 0.051$ (68$\%$ CL). When this is confirmed
in the future by other experiments with even better sensitivities, it will
be good news for near-future experiments such as 
T2K~\cite{T2K_expt}, NO$\nu$A~\cite{Ayres:2004js}, T2HK~\cite{Itow:2001ee},
T2KK~\cite{Hagiwara:2006vn}. Their goal of measuring the CP-violating
effect will be more reachable.  

This is not the end of the story for the neutrino oscillation community.
Another issue is the determination of the sign of $\sin\theta_{13}$.  
As we enter the era of precision neutrino oscillation experiments, terms 
linear in $\sin\theta_{13}$ may no longer be negligible in fitting the 
mixing angles.  
From the Jarlskog invariant quantity~\cite{Jarlskog:1985ht, Wu:1985ea} 
of CP violation, one sees that the sign of $\sin\theta_{13}$
would also have an impact on the $\CPV$ determination.  However, in the 
conventional parametrization, the reactor and solar experiments are only 
sensitive to $\sin^2 \theta_{13}$.  For LBL accelerator or atmospheric 
experiments, the observable has terms linear in $\sin\theta_{13}$, but 
only sub-dominant.  Despite the unfavorable situation one faces when working 
in the conventional parametrization, a first attempt to determine the sign 
of $\sin\theta_{13}$ has been made in Ref.~\cite{Roa:2009wp} using the LBL 
accelerator, CHOOZ, and atmospheric neutrino data.

In fact the conventional parametrization is not the only way to establish 
the mixing matrix.  As proposed by several 
authors~\cite{Schechter:1980gr,Fritzsch:1997st,Zheng:2010kp,Giunti_Kim,
Choubey:2000bf,Huang:2011by}, the corresponding mixing parameters may be 
more accessible, without sacrificing accuracy, in other parametrizations.
In this work we follow the approach of Ref.~\cite{Huang:2011by} to explore 
this possibility.  Besides the conventional parametrization, we perform a 
global neutrino oscillation analysis adopting an alternative parametrization, 
$R_{13} W_{12} R_{23}$, in which the observables have leading terms linear in 
any of the three mixing angles. Equipped with the analysis results obtained in
these two parametrizations, we will be eligible to address a couple of issues. 
Can one determine the sign of $\sin \theta_{13}$?
Are there other parametrizations which can provide comparable sensitivities 
in extracting neutrino oscillation parameters as the conventional one? 
How are the three mixing angles correlated with each other therein?
Do the matter (MSW) effects~\cite{Wolfenstein_msw, Mik_Smir_msw} affect 
the sign of $\sin \theta_{13}$? 

This paper is organized as follows.
In Section~\ref{sec:anal} we briefly describe our data fitting procedure in 
the conventional parametrization (to be denoted by $\RepA$). 
In Section~\ref{sec:anal_results} we present and discuss our results obtained 
from a similar analysis done in an alternative parametrization (to be denoted 
by $\RepD$).  Section~\ref{sec:summary} gives our summary and outlook.  
Individual analysis approaches for each neutrino oscillation experiment 
are detailed in Appendix~\ref{sec:apdx_anal_solar} 
through~\ref{sec:apdx_anal_atmos}. 

%
\section{Analysis}\label{sec:anal}

Table~I in Ref.~\cite{Huang:2011by} indicates that the 
predicted neutrino mixing angles in the parametrizations $\Rbc \Wac \Rab$, 
$\Rbc \Wab \Rac$, and $\Rac \Wbc \Rab$ come up with similar values, while 
those in $\Rac \Wab \Rbc$, $\Rab \Wbc \Rac$, and $\Rab \Wac \Rbc$ end up 
with similar values. We thus preform a global fit of the neutrino mixing 
parameters using parameterizations 
$R_{23}W_{13}R_{12}$ and $R_{13}W_{12}R_{23}$, taking advantage of the 
fact that these two parameterizations will have dissimilar outcomes.
Following Ref.~\cite{Huang:2011by}, we denote them as the $\RepA$ and $\RepD$ 
'representations' respectively, where representation $\RepA$ corresponds to 
the conventional mixing matrix parametrization.  In what follows, notations 
($\DAmij$, $\thA_{ij}$) and ($\DDmij$, $\thD_{ij}$) stand for the oscillation 
parameters directly determined using representations $\RepA$ and $\RepD$, 
respectively. On the other hand, the notation $\thDA_{ij}$ symbolizes the 
three mixing angles that are initially extracted from representation $\RepD$  
and then translated to representation $\RepA$ by the transformation method 
described in Ref.~\cite{Huang:2011by}. The transformation from representation
$\RepD$ to $\RepA$ is outlined in Appendix~\ref{apdx:Gfit_NumParamSol_DA}.

The global neutrino data used in this work include 
{\it (i)} solar data from rates measured in chlorine~\cite{Cleveland:1998nv} 
and gallium~\cite{Abdurashitov:2009tn} experiments, the rate of $^{7}$Be 
solar neutrinos measured in the Borexino~\cite{Arpesella:2008mt} experiment, 
Super-Kamiokande (SK) phase I \& II day/night 
spectra~\cite{Hosaka:2005um,Cravens:2008zn}, 
and SNO phases I \& II $\nu_e$ survival probability~\cite{Aharmim:2009gd};
{\it (ii)} reactor data from KamLAND~\cite{Gando:2010aa} and 
CHOOZ~\cite{Apollonio:2002gd};
{\it (iii)} LBL accelerator data from K2K~\cite{Ahn:2006zza} and MINOS 
$\nu_{\mu}$ disappearance channel~\cite{Adamson:2011ig} and $\nu_e$ 
appearance channel~\cite{MINOS_nue_app_2011}; and 
{\it (iv)} atmospheric data from SK phase I~\cite{Hosaka:2006zd} and 
SNO~\cite{Aharmim:2009zm}.

In this work, the re-evaluated $\barnue$ flux from nuclear power plants
\cite{Mueller:2011nm} is not taken into consideration for the reactor data. 
In addition, the large number of bins in the atmospheric data of SK phases 
II and III, as well as the lack of information, prevents us from reproducing 
their results.  Therefore we do not include the SK-II and SK-III atmospheric 
data in our work.  Since our purpose is to investigate the neutrino mixing 
phenomenology in different parametrizations using the same data sets and 
analysis conditions, the absence of these factors should not have any impact 
on the conclusions of this work.  Described below is our analysis of the 
global neutrino data.

Appendix~\ref{sec:apdx_anal_solar} outlines the analyses for each solar 
experiment employed in this work which includes chlorine, gallium, Borexino, 
SK and SNO. We use the Bahcall solar neutrino 
spectra~\cite{Solar_Nu_Eng_Bahcall} except for the $^{8}$B neutrino spectrum, 
which is from Ref.~\cite{Winter:2004kf}. Neutrino survival probabilities 
in the Sun are estimated using the BS05(OP) model~\cite{Bahcall:2004pz} for 
the neutrino production rates at different solar radii. We do not use the more 
recent BPS09(GS) or BPS09(AGSS09) models~\cite{Serenelli:2009yc} because of 
the conservative model uncertainties. For estimating neutrino survival 
probabilities inside the Earth, the Earth density profile of 
PEM-C~\cite{PEM-C} (rather than PREM~\cite{PREM}) is adopted in this work. 
It assumes the continental crust for the outer most layer of the Earth 
where the solar neutrinos enter the detectors. In our analysis, the flux of
$^{8}$B solar neutrinos is a nuisance parameter while the fluxes of other 
solar neutrinos are taken from the BS05 model~\cite{Bahcall:2004pz}. 
Neutrino oscillation probabilities are calculated using the adiabatic 
approximation~\cite{Ioannisian_2004}. It has been verified to yield 
equivalent results as the numerical calculation does for several sets 
of neutrino oscillation parameters in the LMA region. 

Our analyses of oscillation parameter fitting using the KamLAND and the 
CHOOZ $\barnue$ oscillation events are delineated in 
Appendix~\ref{sec:apdx_anal_reactor}.  Owing to the short distances between 
the source and the detector, matter effects are generally insignificant and
oscillation probabilities in vacuum suffice. This is also the case for 
accelerator neutrino analyses.  Appendix~\ref{sec:apdx_anal_accel} describes 
the analyses of oscillation parameter fitting using the K2K and the MINOS 
$\numu$ disappearance channel as well as the MINOS $\nue$ appearance channel.

Atmospheric neutrino data from the SK~\cite{Hosaka:2006zd} and the 
SNO~\cite{Aharmim:2009zm} experiments are included in our analysis as 
described in Appendix~\ref{sec:apdx_anal_atmos}. We first employ the NUANCE 
package~\cite{Casper:2002sd} to simulate atmospheric neutrino events assuming 
no oscillation effects. Neutrino oscillations in the atmosphere and inside 
the Earth are then incorporated by the "weighting factors", 
Eq.~(\ref{eq:weighting_factors}), with the matter effects taken into account 
by following the prescription of Ref.~\cite{Barger:1980tf}. We apply similar 
criteria and cuts on the kinematics of the simulated events so that we achieve 
the same selection efficiency as Ref.~\cite{Hosaka:2006zd} does for SK 
atmospheric data and as Ref.~\cite{Aharmim:2009zm} does for SNO atmospheric
data.

\begin{figure}
\centerline{\psfig{figure=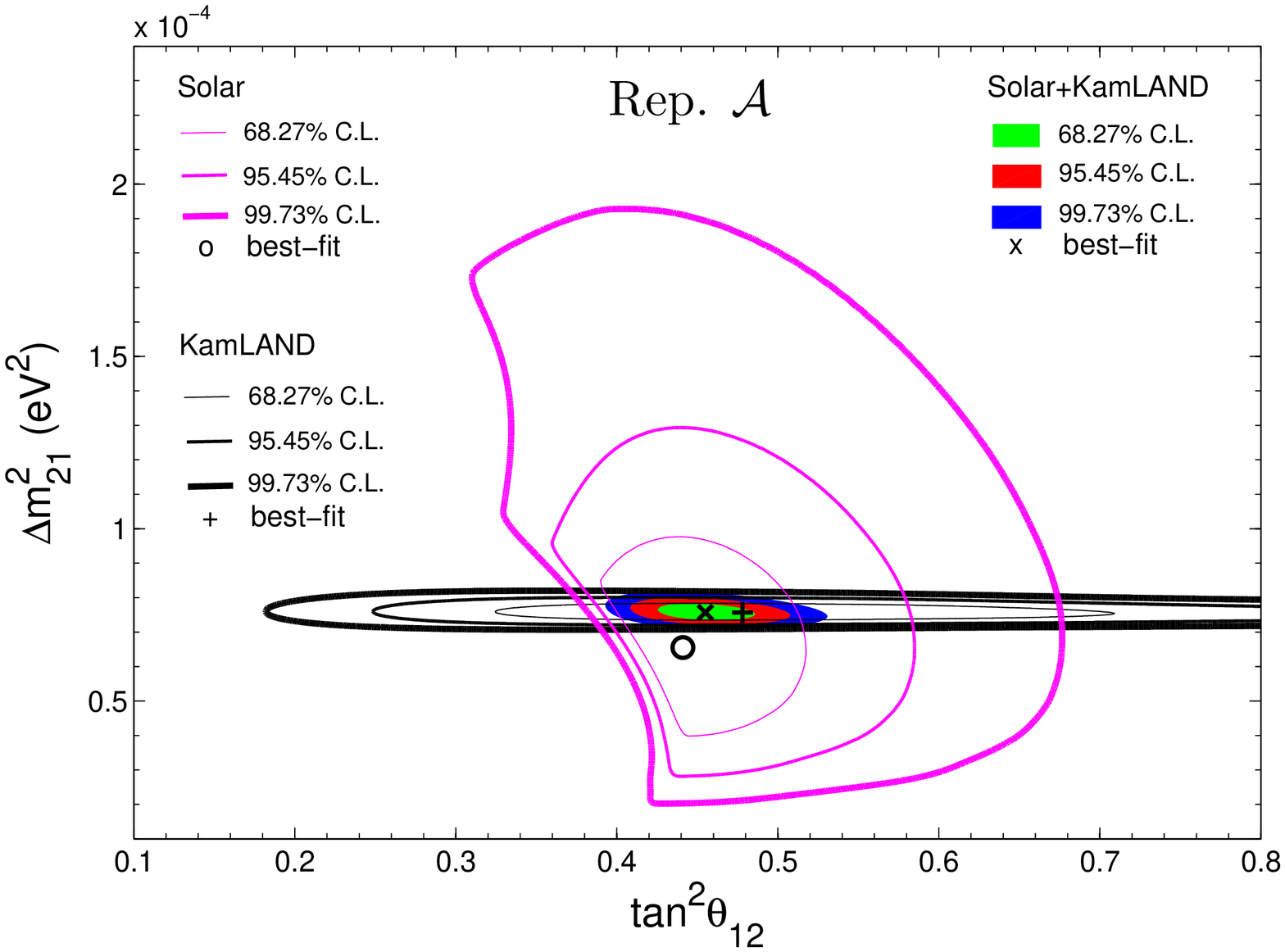,width=87mm}}
\centerline{\psfig{figure=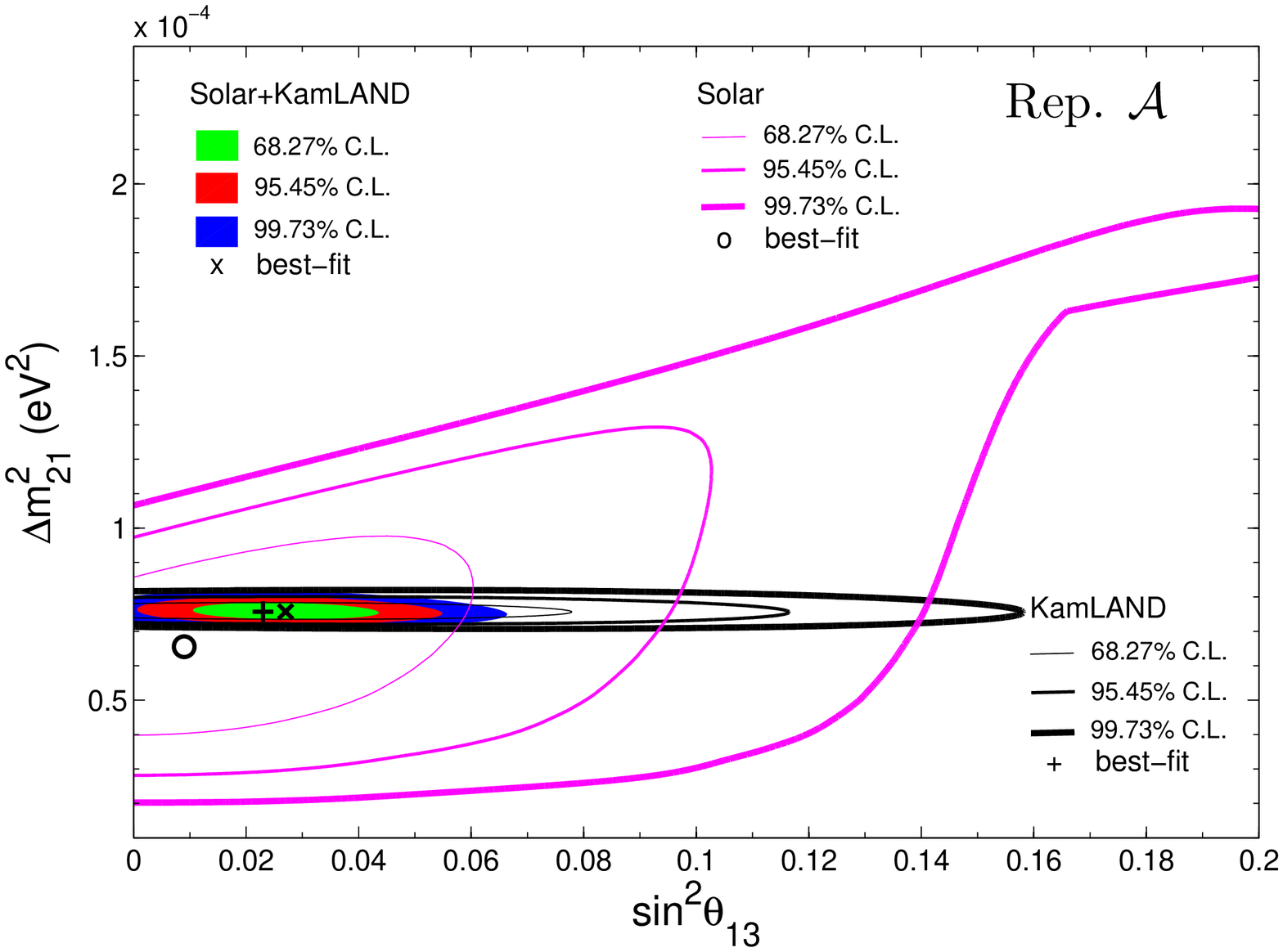,width=87mm}}
\centerline{\psfig{figure=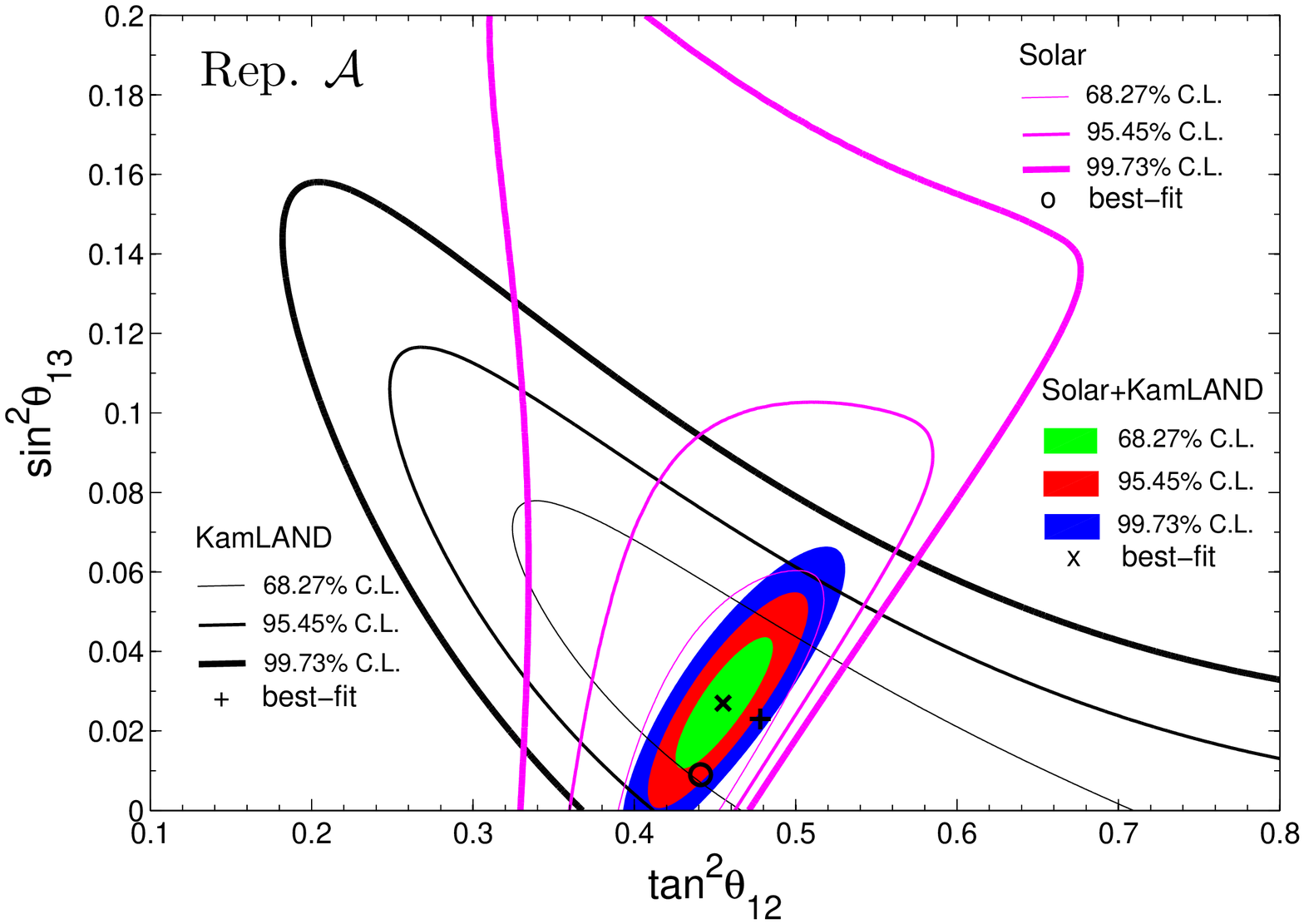,width=87mm}}
\caption{Results of the solar and KamLAND oscillation parameter analysis
         in a three-flavor hypothesis performed in representation $\RepA$,
         where $\DAmbc = 2.4 \times10^{-3}$ eV$^2$ and $\sinthacA = 0.5$
         are assumed.}
\label{fig:SolKL_caseA}
\end{figure}

%
%
\begin{table*}[th]
\begin{tabular}{cccccc}  \hline\hline
Data Sample  &  $|{\DAmab}|$         &  $|\DAmbc|$
             &  $\thA_{12}$  &  $\thA_{23}$  & \vspace{0.1cm} $|\thA_{13}|$ \\
             &  [$10^{-5}$ eV$^2$]  &  [$10^{-3}$ eV$^2$]
             &  [$\tanthabA$]   &  [$\sinthbcA$]
             & \vspace{0.1cm} [$\sinthacA$]
                                            \vspace{0.1cm}  \\ \hline
Solar-only   & $6.55^{+3.21}_{-2.57}$      &
             & $33.587^{+2.156}_{-1.603}$  &
             & \vspace{0.1cm} $5.444^{+8.735}_{-5.444}$     \\
             &                             &
             & [$0.441^{+0.077}_{-0.051}$] &
             & \vspace{0.1cm} [$0.009^{+0.051}_{-0.009}$]
                                            \vspace{0.1cm}  \\ \hline
KamLAND-only & $7.57^{+0.27}_{-0.22}$      &
             & $34.659^{+5.459}_{-5.010}$  &
             & \vspace{0.1cm} $8.723^{+7.495}_{-8.723}$     \\
             &                             &
             & [$0.478^{+0.232}_{-0.154}$] &
             & \vspace{0.1cm} [$0.023^{+0.055}_{-0.023}$]
                                            \vspace{0.1cm}  \\ \hline
Solar+KamLAND& $7.57^{+0.26}_{-0.22}$      &
             & $34.001^{+0.881}_{-0.900}$  &
             & \vspace{0.1cm} $9.458^{+2.651}_{-3.718}$     \\
             &                             &
             & [$0.455^{+0.031}_{-0.030}$] &
             & \vspace{0.1cm} [$0.027^{+0.020}_{-0.017}$]
                                            \vspace{0.1cm}  \\ \hline
Accel+Atmos  &  & $2.30^{+0.21 }_{-0.11 }$
             &  & $43.739^{+4.391}_{-4.391}$
             & \vspace{0.1cm}  $5.132^{+3.591}_{-5.132}$    \\
 +CHOOZ      &  &
             &  & [$0.478^{+0.058}_{-0.058}$]
             & \vspace{0.1cm} [$0.008^{+0.015}_{-0.008}$]
                                            \vspace{0.1cm}  \\ \hline
Global       &   &   &   &
             & \vspace{0.1cm} $7.492^{+1.231}_{-1.753}$     \\
$(\Delta \chi^2 = 1.0)$
             &   &   &   &
             & \vspace{0.1cm} [$0.017^{+0.006}_{-0.007}$]   \\ \hline
Global       &   &   &   &
             & \vspace{0.1cm} $7.492^{+1.787}_{-3.049}$     \\
$(\Delta \chi^2 = 2.30)$
             &   &   &   &
             & \vspace{0.1cm} [$0.017^{+0.009}_{-0.011}$]   \\ \hline \hline
\end{tabular}
\caption{Summary of our best-fit oscillation parameters, $\DAmij$ and
         $\thA_{ij}$, in representation $\RepA$, where the mxing angles,
         $\thA_{ij}$, are in unit of degrees. In our analysis we only
         fitted $\sinthbcA$ in the range below 0.5. In order to compare
         the 68$\%$ CL constraints determined in representations $\RepA$
         and $\RepD$, we assume the upper and lower bounds from
         accelerator$+$CHOOZ$+$atmospheric data are about the same.}
\label{table:best_fits_A}
\end{table*}
We first perform a global analysis using representation $\RepA$ in order 
to compare with existing results from the literature. As the current global 
neutrino data is insensitive to $\CPV$, we have selected $\CPV=0$.
According to the study presented in Ref.~\cite{Huang:2011by}, this choice 
also turns out to be adequate for representation $\RepD$ since a rephasing 
still gives $\CPV=0$ therein.
Table~\ref{table:best_fits_A} summarizes our best-fit results for the  
oscillation parameters from various three-flavor analyses in this case. 
Assuming CPT invariance, the results from the KamLAND reactor experiment and
all solar experiments are combined in the $\chi^2$ function:
$\chi^2_{\rm sol + KL} \equiv \chi^2_{\rm solar} + \chi^2_{\rm KL}$.
We minimize $\chi^2_{\rm sol + KL}$ with respect to the oscillation parameters
$\DAmab$, $\tanthabA$, and $\sinthacA$, as well as the $^{8}$B solar neutrino 
flux in a three-flavor hypothesis.  The other two parameters are fixed at
$\DAmbc = 2.4 \times10^{-3}$ eV$^2$ and $\sinthbcA = 0.5$. 
The best-fit values are found to agree with the results presented in 
Refs.~\cite{Aharmim:2009gd,Gando:2010aa}. 
Figure~\ref{fig:SolKL_caseA} shows the allowed regions in the planes of 
($\tanthabA$, $\DAmab$), ($\sinthacA$, $\DAmab$), and 
($\tanthabA$, $\sinthacA$).

\begin{figure}
\centerline{\psfig{figure=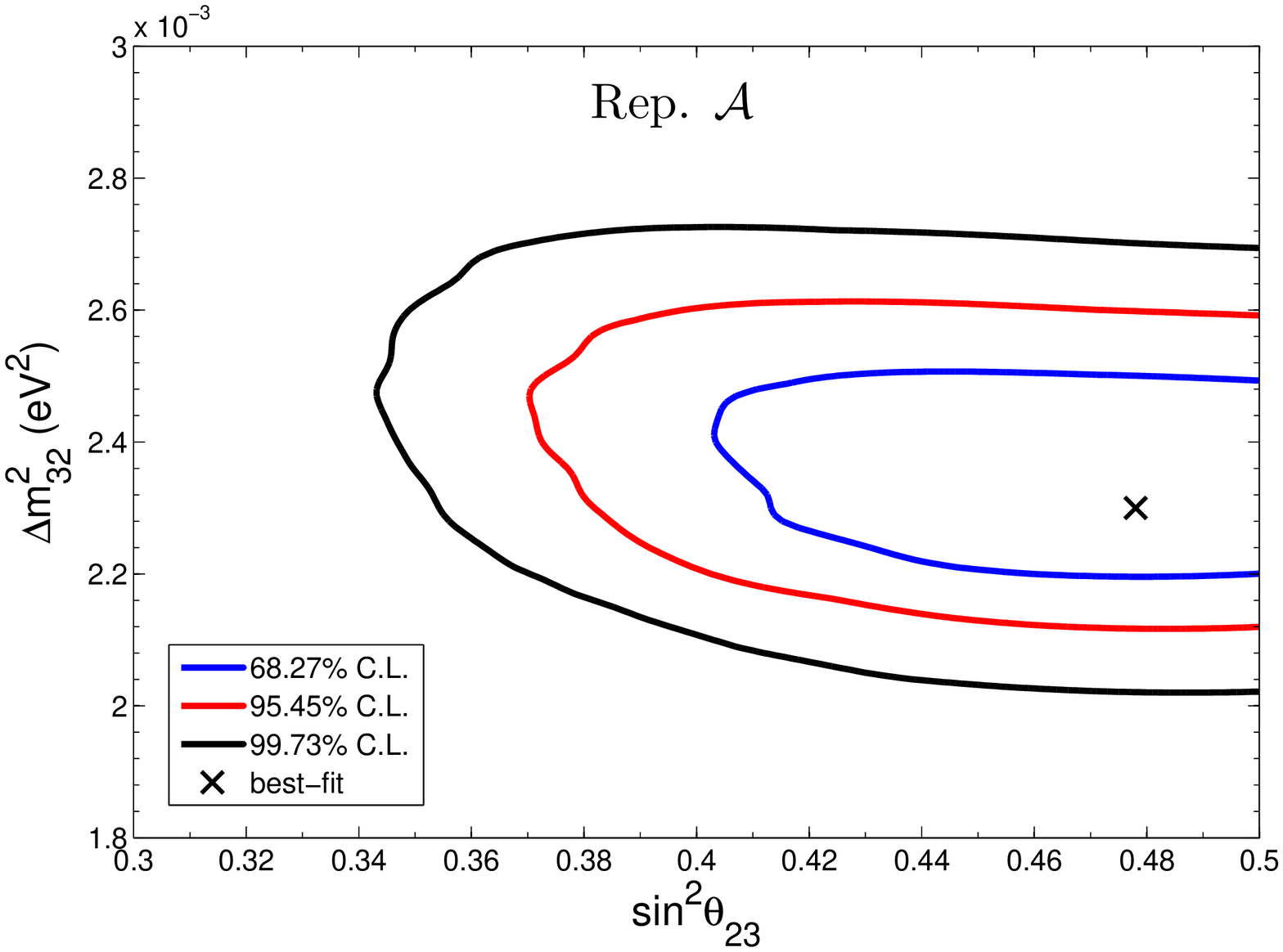,width=87mm}}
\centerline{\psfig{figure=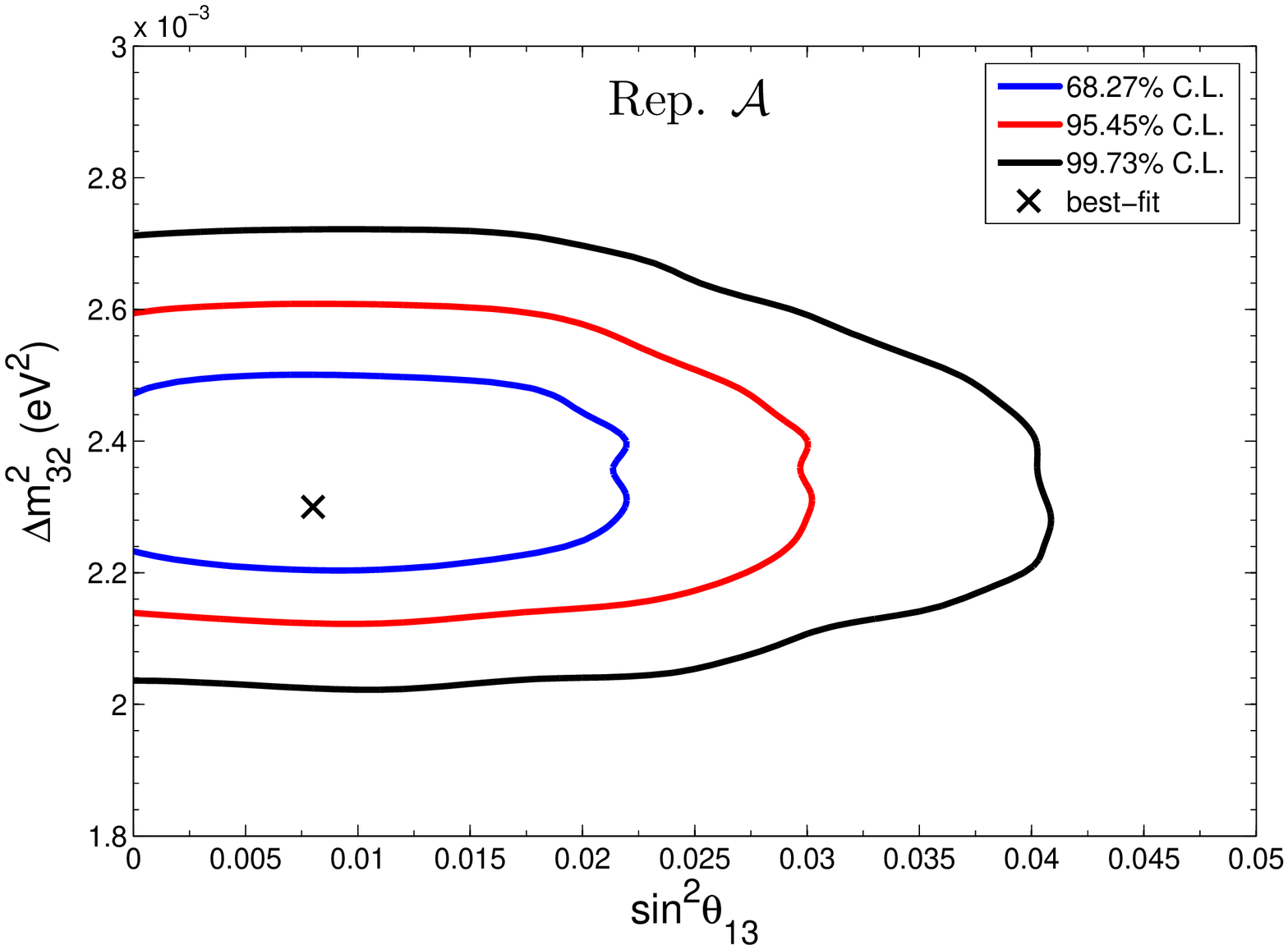,width=87mm}}
\centerline{\psfig{figure=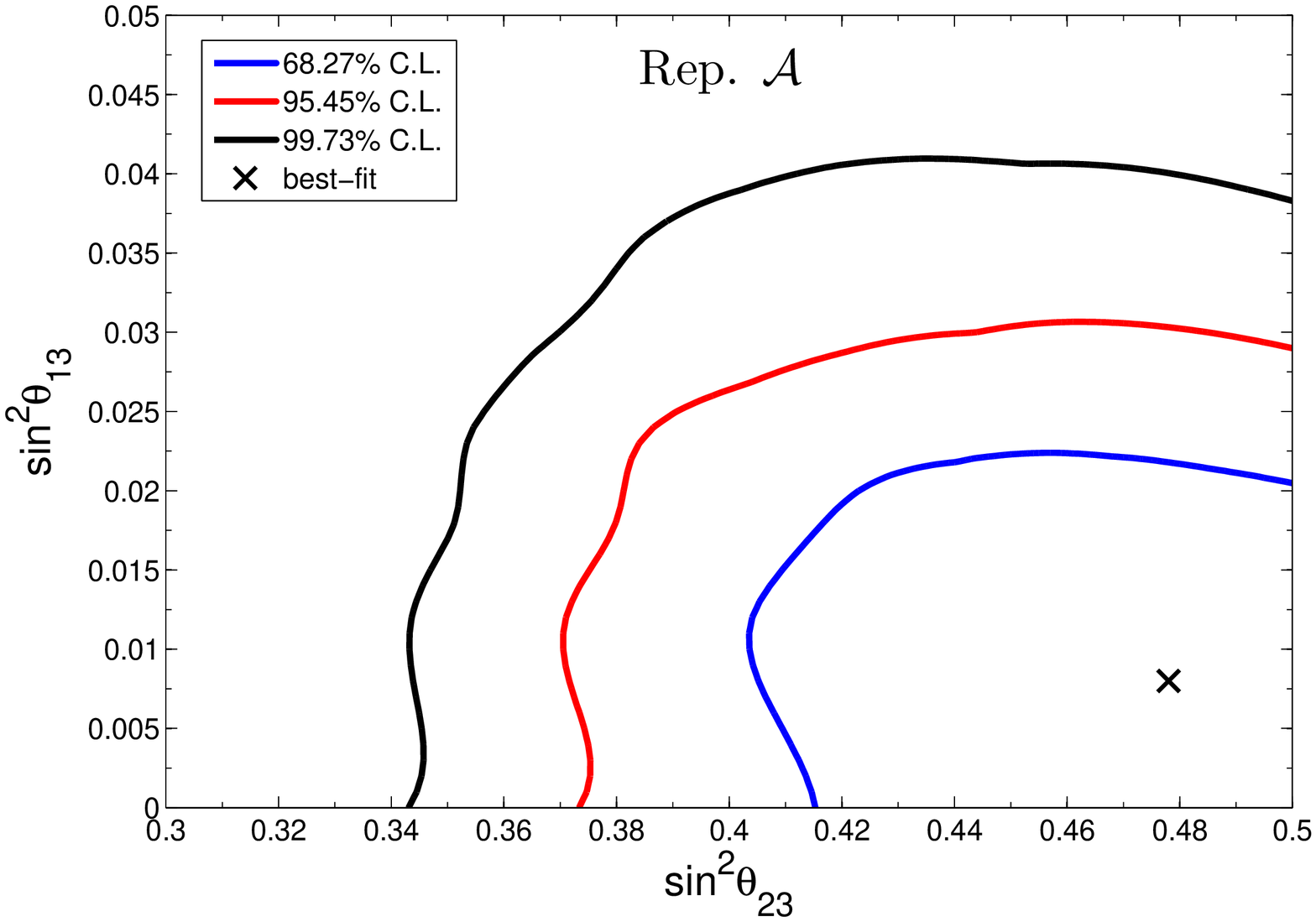,width=87mm}}
\caption{Three-flavor oscillation parameter results in representation 
         $\RepA$ for the combined analysis of the LBL accelerator, 
         CHOOZ, and atmospheric data sets, where the parameters 
         $\DAmab = 7.67 \times10^{-5}$ eV$^2$ and $\tanthabA = 0.427$ 
         are assumed.}
\label{fig:numu_caseA}
\end{figure}

The $\chi^2$ functions for K2K, MINOS $\numu$ disappearance, MINOS 
$\nue$ appearance, CHOOZ, as well as SK and SNO atmospheric neutrino data 
are also summed together: 
$\chi^2_{\rm acc + CHOOZ + atm} \equiv \chi^2_{\rm acc} + \chi^2_{\rm CHOOZ} 
+ \chi^2_{\rm atm}$.  
By minimizing this $\chi^2$ with respect to the oscillation parameters 
$\DAmbc$, $\sinthbcA$, and $\sinthacA$ while fixing 
$\DAmab = 7.67 \times10^{-5}$ eV$^2$ and $\tanthabA = 0.427$, and assuming 
a normal mass hierarchy, the best-fit values are found to be consistent 
with the most recent results in Ref.~\cite{Schwetz:2011qt, Fogli:2011qn}.
Figure~\ref{fig:numu_caseA} shows the likelihood contours in the planes of 
($\sinthbcA$, $\DAmbc$), ($\sinthacA$, $\DAmbc$), and 
($\sinthbcA$, $\sinthacA$).

\begin{figure}
\centerline{\psfig{figure=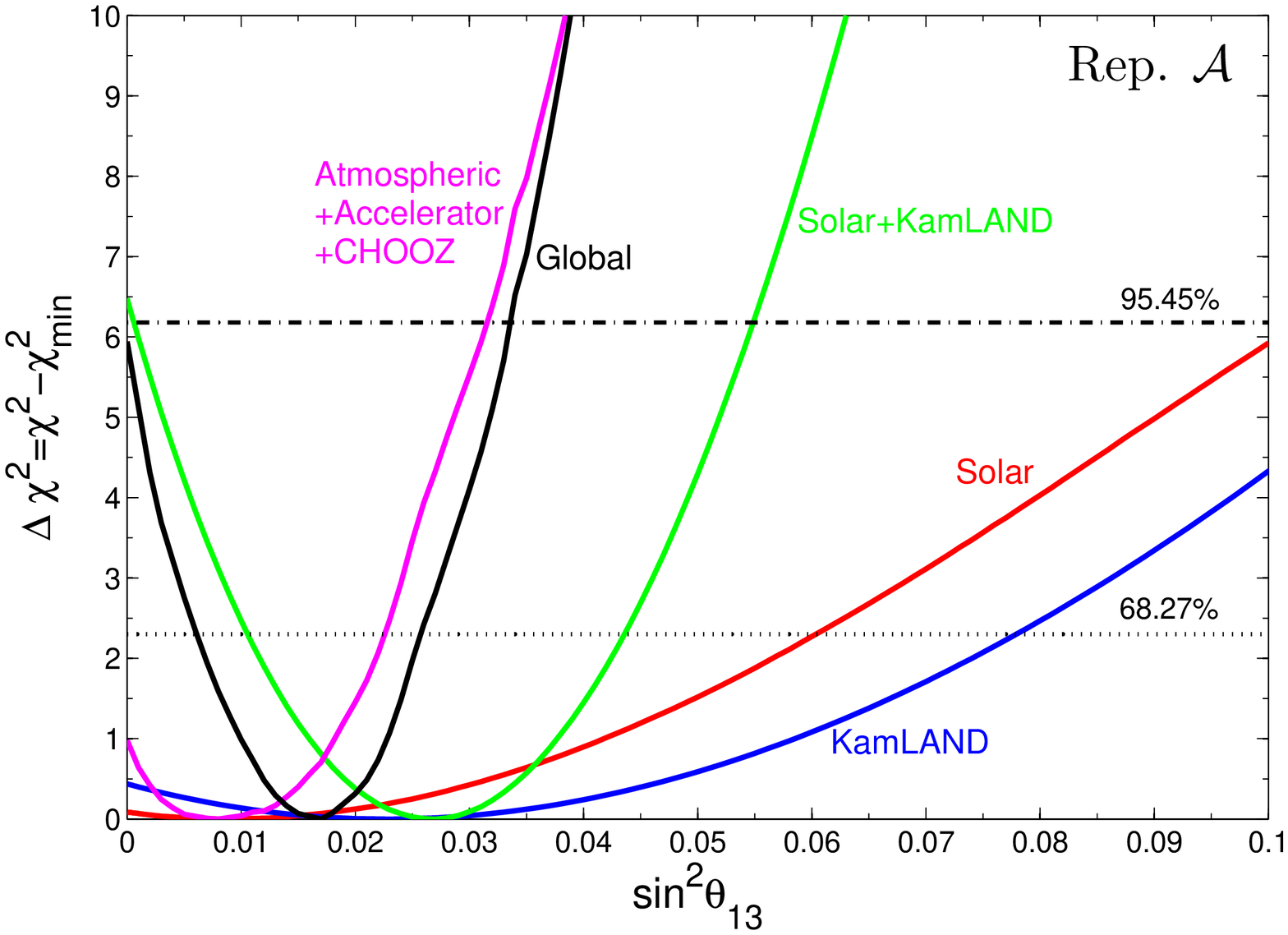,width=87mm}}
\caption{$\Delta \chi^2$ as a function of $\sinthacA$ for various
         combinations of oscillation data where the undisplayed 
         parameters are floated.} 
\label{fig:Dchi2_caseA}
\end{figure}

Finally, for individual data sample mentioned above, i.e. the solar-only, 
KamLAND, solar$+$KamLAND, and accelerator$+$CHOOZ$+$atmospheric, 
the differences ($\Delta \chi^2$) between $\chi^2$ and the minimum 
$\chi^2_{\rm min}$ as a function of $\sinthacA$ are obtained by 
marginalizing the other two oscillation parameters.  Our results are
shown in Fig.~\ref{fig:Dchi2_caseA} for 
$\Delta \chi^2_{\rm solar}$, $\Delta \chi^2_{\rm KL}$, 
$\Delta \chi^2_{\rm sol + KL}$ and $\Delta \chi^2_{\rm acc + CHOOZ + atm}$. 
The global result is then 
\begin{equation}
  \Delta \chi^2_{\rm global} (\sinthacA) \equiv \Delta \chi^2_{\rm sol + KL}
  + \Delta \chi^2_{\rm acc + CHOOZ + atm}\, .
\end{equation}
Our result for $\sinthacA$ is consistent with those from other global neutrino 
data analyses 
(e.g., \cite{Aharmim:2009gd,Gando:2010aa,Fogli:2011qn,Schwetz:2011qt,
GonzalezGarcia:2010er}).
We also find a weak hint for non-zero $\sinthacA$ at 95$\%$ CL, 
but it is not easy to determine the sign of $\sin \thA_{13}$.


%
\section{Results and Discussions}\label{sec:anal_results}

Changing to representation $\RepD$ and fixing $\CPV=0$, we perform a similar 
$\chi^2$ analysis as that in previous section using exactly the same data 
sets.  After determining the oscillation parameters $\DDmij$ and $\thD_{ij}$, 
with the transformation strategy described in 
Appendix~\ref{apdx:Gfit_NumParamSol_DA}, we also perform a translation 
of the best-fit mixing angles back to representation $\RepA$, denoted by 
$\thDA_{ij}$.  Both results are given in Table~\ref{table:best_fits_DA} 
for a comparison with the corresponding $\DAmij$ and $\thA_{ij}$ retrieved 
directly in representation $\RepA$.  
\begin{table*}[th]
\begin{tabular}{ccccccccc}  \hline\hline
Data Sample  &  $|{\DDmab}|$         &  $|\DDmbc|$
             &  $\thD_{12}$  &  $\thD_{23}$  & \vspace{0.1cm} $\thD_{13}$
             &  $\thD_{12}$  &  $\thD_{23}$  & \vspace{0.1cm} $\thD_{13}$ \\
             &  [$10^{-5}$ eV$^2$]  &  [$10^{-3}$ eV$^2$]
             & [$\thDA_{12}$ & $\thDA_{23}$  & \vspace{0.1cm} $\thDA_{13}$]
             & [$\thDA_{12}$ & $\thDA_{23}$  & \vspace{0.1cm} $\thDA_{13}$] \\
             &                      &
             & \multicolumn{3}{c}{\emph{Global Minimum}}
             & \multicolumn{3}{c}{\emph{Second Minimum}}
                                            \vspace{0.1cm}  \\ \hline
Solar-only   & $6.00^{+3.20}_{-2.60}$
                     &       &  25.0   &  57.0   &  \vspace{0.1cm} -24.0
                             &         &         &         \\
             &       &       & [33.662 & 49.827  & \vspace{0.1cm} 5.870]
                             &         &         &         \\ \hline
KamLAND-only & $7.60^{+0.23}_{-0.25}$
                     &       &  36.0   &  50.0   & \vspace{0.1cm} -46.0
                             &         &         &         \\
             &       &       & [55.362 & 38.814  & \vspace{0.1cm} -8.604]
                             &         &         &         \\ \hline
Solar+KamLAND& $7.60^{+0.23}_{-0.25}$
                     &       &  11.0   &  57.0   & \vspace{0.1cm} -34.0
                             &         &         &         \\
             &       &       & [34.300 & 56.689  & \vspace{0.1cm} -9.898]
                                                           \\ \hline
Accel+Atmos  &       & $2.40^{+0.18}_{-0.12}$
                             &  27.0   &  50.0   & \vspace{0.1cm} -38.0
                             &         &         &         \\
 +CHOOZ      &       &       & [44.978 & 43.445  & \vspace{0.1cm} -6.990]
                             &         &         &         \\ \hline
Global       &  7.6  &  2.4  & $19.0^{+1.8}_{-3.2}$
             & $46.0^{+4.6}_{-4.0}$   &  $-29.0^{+1.5}_{-2.5}$
             & $28.5^{+2.0}_{-2.5}$    &  $51.5^{+3.5}_{-4.5}$
             & \vspace{0.1cm} $-20.0^{+2.5}_{-2.0}$          \\
$(\Delta \chi^2 = 2.30)$
             &       &       & [33.461 &  43.326 &  $-7.582$]
                             & [33.509 &  43.980 &    7.932]  \\
             &       &       & [$\pm$0.891  &  $\pm$4.391  &  $\pm$2.418]
                             & [$\pm$1.760  &  $\pm$4.118  &  $\pm$2.502]
                                                     \\  \hline  \hline
\end{tabular}
\caption{Summary of our best-fit oscillation parameters, $\DDmij$ and
         $\thD_{ij}$, in representation $\RepD$, and the translated mixing
         angles, $\thDA_{ij}$ in representation $\RepA$, where all of the
         mixing angles are in unit of degrees. Note that the 68$\%$ CL
         constraints of the best-fit results from the second minimum
         is taken at $\Delta \chi^2 = \chi^2 - \chi^2_{min} = 2.3$ with
         $\chi^2_{min}$ being the $\chi^2$ value at the second minimum.
         The uncertainties of $\thDA_{ij}$ are listed at the last row.}
\label{table:best_fits_DA}
\end{table*}
%

We found that in representation $\RepD$, individual data sets, i.e. the 
global solar data, \mbox{KamLAND}, \mbox{solar$+$KamLAND}, and the combined 
\mbox{accelerator$+$CHOOZ$+$atmospheric} data, cannot constrain the three 
mixing angles well.  However, when combined altogether, they infer bounds 
at 68$\%$ confidence level (CL) on the three mixing angles $\thD_{ij}$ which 
are comparable to those on $\thA_{ij}$ (cf. Table~\ref{table:best_fits_DA}),
except for $\thD_{12}$.  We therefore give only the bounds obtained from the 
combined global neutrino data.

Using the global solar data, we find that the best-fit values of $\thDA_{12}$ 
and $\thDA_{13}$ are consistent with those of $\thA_{12}$ and $\thA_{13}$, 
respectively. In representation $\RepA$, the solar data are not 
sensitive to $\thA_{23}$, thus one should not naively compare the 
translated best-fit result of $\thDA_{23} \sim 50^{\circ}$ to the input 
value of $\thA_{23} = 45^{\circ}$.  The KamLAND data are insensitive to 
$\thA_{23}$ too and have less constraining power on $\thA_{12}$. These two 
facts explain why the translated values of $\thDA_{12}$ and $\thDA_{23}$  
are very different from the expected values, $\thA_{12}$ and $\thA_{23}$.  
Surprisingly, the translated best-fit $\thDA_{13}$ turns out to be almost 
the same as $\thA_{13}$ in magnitude, but with an opposite sign.  
Like the KamLAND-only results, the combined solar and KamLAND results
produce best-fit values where 
$\thDA_{12} \sim \thA_{12}$ and $|\thDA_{13}| \sim |\thA_{13}|$, 
and $\thDA_{13} < 0$. 

Similarly, combining the LBL accelerator, CHOOZ, and atmospheric neutrino 
data, the translated best-fit $\thDA_{23}$ and $|\thDA_{13}|$ are 
consistent with the best-fit $\thA_{23}$ and $|\thA_{13}|$, respectively.  
The combined \mbox{accelerator$+$CHOOZ$+$atmospheric} data are not sensitive 
to $\thA_{12}$, so a comparison of the best-fit value of 
$\thDA_{12} \sim 45^{\circ}$ with the input value of $\thA_{12} = 33.2^\circ$ 
is not necessary. Interestingly for this case, it was found that 
the best fit $\thDA_{13} < 0$ as well.  This finding has been 
discussed in Ref.~\cite{Roa:2009wp} using a similar data sample.

\begin{figure}
\centerline{\psfig{figure=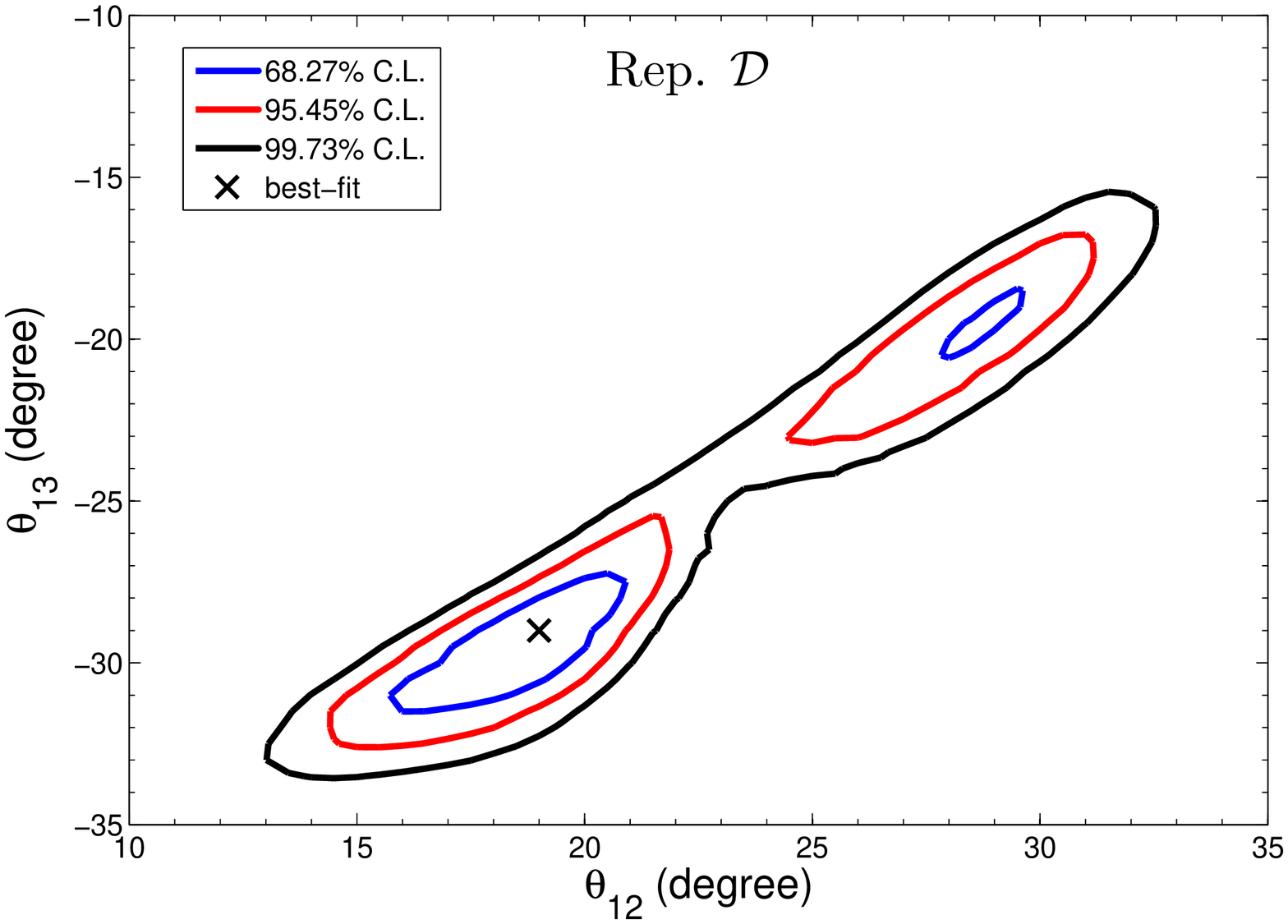,width=87mm}}
\centerline{\psfig{figure=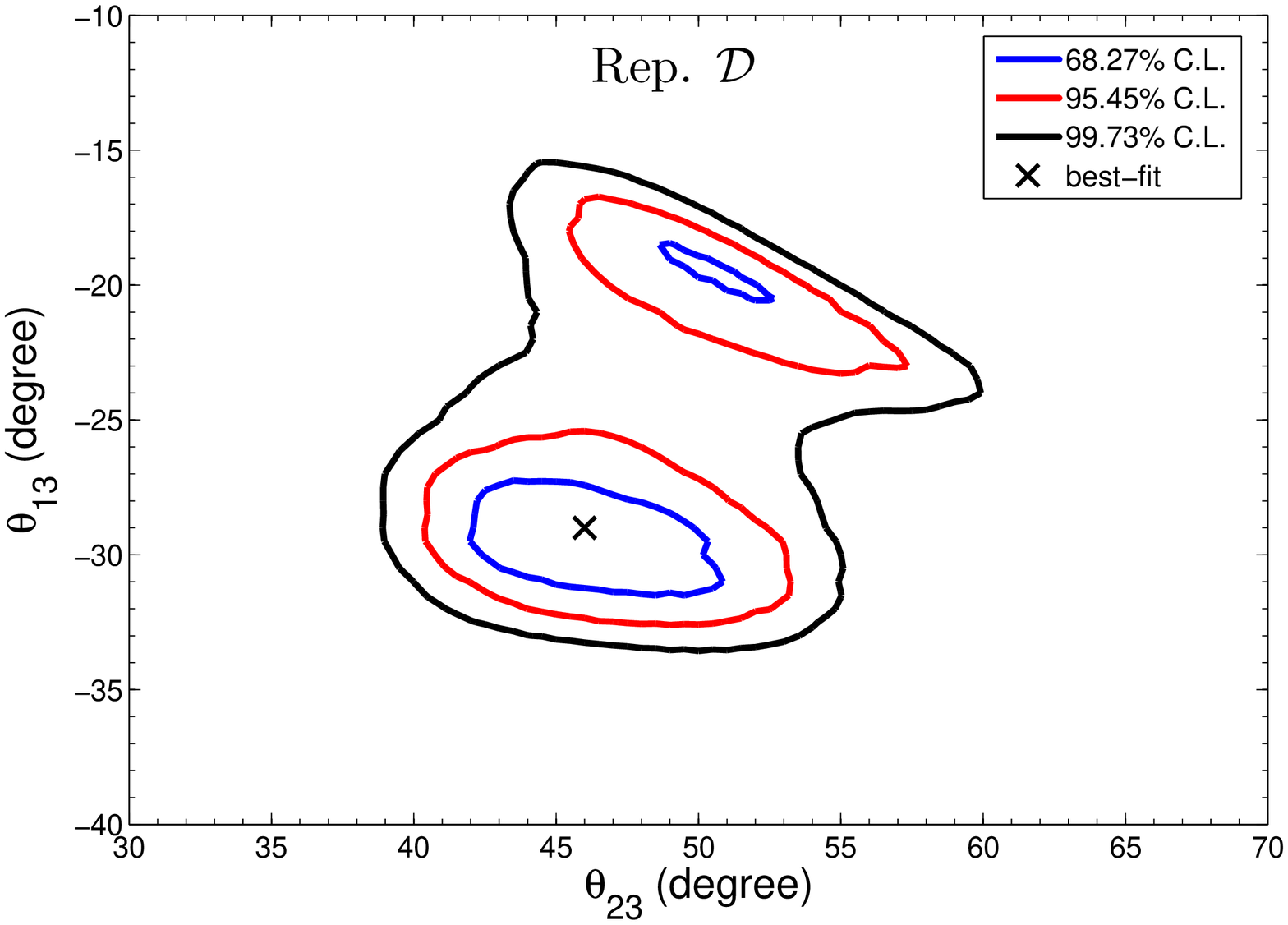,width=87mm}}
\centerline{\psfig{figure=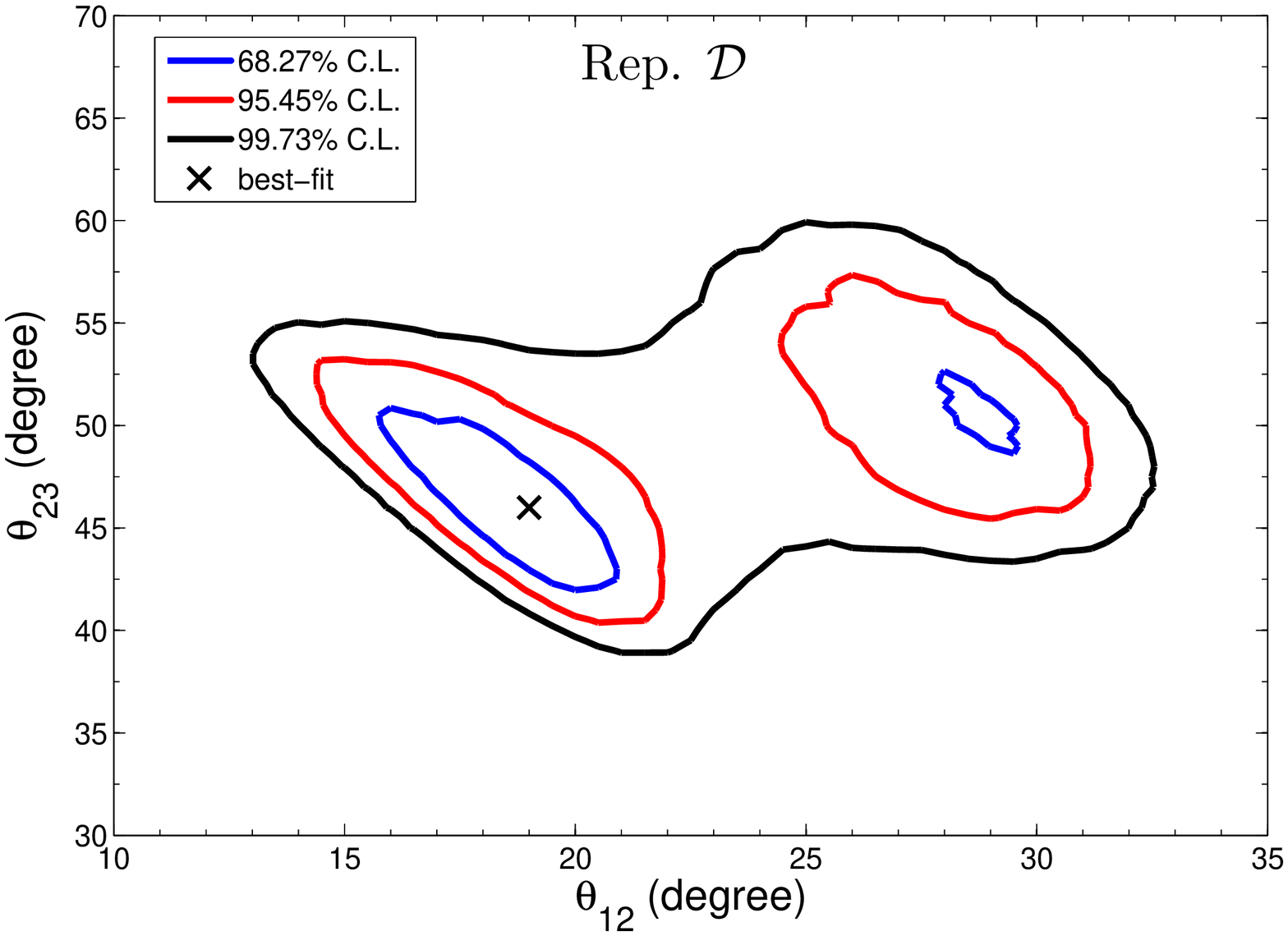,width=87mm}}
\caption{Three-flavor hypothesis mixing angle results for the combined 
         analysis of global neutrino data in representation $\RepD$.  
         The neutrino mass differences are fixed to the values
	     $\DDmab = 7.6 \times10^{-5}$ eV$^2$ and
         $\DDmbc = 2.4 \times10^{-3}$ eV$^2$, and the normal mass 
         hierarchy is assumed.}
\label{fig:Global_CaseD_deg}
\end{figure}

In spite of the poor constraints on $\thD_{ij}$, the parameter $\DDmab$ 
determined by analyzing the solar, KamLAND and solar$+$KamLAND data,
as well as $\DDmbc$ from the combined analysis of the LBL accelerator, 
CHOOZ and the atmospheric data agree well with those directly extracted 
in representation $\RepA$.  Equipped with the above individual $\chi^2$ 
calculation, we perform a global analysis straightforwardly via
\begin{equation}
  \Delta \chi^2_{\rm global} (\thD_{12}, \thD_{23}, \thD_{13}) \equiv 
     \Delta \chi^2_{\rm sol + KL} + \Delta \chi^2_{\rm acc + CHOOZ + atm}\, .
\end{equation}
Here we fix $\DDmab = 7.6 \times 10^{-5}$ eV$^{2}$ and 
$\DDmbc = 2.4 \times 10^{-3}$ eV$^{2}$,
the best-fit values determined from the \mbox{solar$+$KamLAND} and the 
\mbox{acc$+$CHOOZ$+$atm} analysis, respectively. 
Our results at 68$\%$ CL are given in Table~\ref{table:best_fits_DA}, 
and Fig.~\ref{fig:Global_CaseD_deg} shows the projected likelihood contour 
plots in the planes of $(\thD_{12}, \thD_{13})$, $(\thD_{23}, \thD_{13})$, 
and $(\thD_{12}, \thD_{23})$.  As already advertised in the beginning of 
this section, we find that the 68$\%$ CL constraints on $\thD_{ij}$ are 
comparable to those on $\thA_{ij}$, with the exception of $\thD_{12}$.

\subsection*{Sign of $\sin \thA_{13}$ }

As we enter the era of precision neutrino oscillation experiments, terms 
linear in $\sin \theta_{13}$ can no longer be neglected.  As pointed out 
in Ref.~\cite{Roa:2009wp}, it will be necessary to perform a parameter 
fitting in the $\sin \thA_{13}<0$ regime as well.
The sign of $\sin \theta_{13}$ also plays the decisive role to the $\CPV$ 
determination.  This can be seen in the Jarlskog invariant quantity of CP 
violation~\cite{Jarlskog:1985ht, Wu:1985ea}. For any representation $\RepH$, 
it is defined as
\begin{equation}
{\mathcal{J}} = \sin 2\thH_{12} \sin 2\thH_{23} \sin 2\thH_{13}
                \cos\thH_{ij} \sin\CPVH\, ,
\label{eq:jarlskog_inv}
\end{equation}  
where $\thH_{ij}$ is the mixing angle seated in the middle of the three 
rotation matrices in the mixing matrix. For representation $\RepH = \RepA$ 
or $\RepD$, both $\sin 2\thH_{12}$ and $\sin 2\thH_{23}$ are positive. 
However, for the existing global neutrino data, the leading terms of the 
observables in representation $\RepA$ are linear in $\sin^2 \theta_{13}$
while the terms linear in $\sin \theta_{13}$ are in the sub-leading terms.
This makes difficult to determine the sign of $\sin \theta_{13}$ in 
representation $\RepA$.

An interesting feature is seen when analyzing neutrino data in representation 
$\RepD$.  Two local minima exist in the global $\chi^2$, one corresponding to 
$\thDA_{13} < 0$ (the global minimum) and the other one $\thDA_{13} > 0$ 
(the second minimum). Apart from this, both minima have very similar values 
of $\thDA_{12}$, $\thDA_{23}$, and $|\thDA_{13}|$, which also agree well with 
the best-fit $\thA_{12}$, $\thA_{23}$, and $|\thA_{13}|$ 
(cf. Table~\ref{table:best_fits_A}), respectively.  This suggests that by 
doing global neutrino data analysis in representation $\RepD$, one has a 
chance to identify the sign of $\sin \thDA_{13}$. In the current case, the 
$\chi^2$ values for the global minimum and the second minimum differ by 
$\sim$2. While this preference for negative $\sin \theta_{13}$ is weak,
the inclusion of data from the current and the near-future neutrino  
experiments may help strengthen the evidence. 

Compared among the various data samples used, the global solar data point 
to  $\thDA_{13} > 0$ while all the other data $\thDA_{13} < 0$.  We do not 
have the explanation for this difference yet.  One may investigate whether 
the matter (MSW) effects~\cite{Wolfenstein_msw, Mik_Smir_msw} influence the 
sign of $\thDA_{13}$ during the propagation of solar neutrinos through the 
Sun and the Earth.
Atmospheric neutrinos propagating through the Earth are also subject to the 
MSW effects (see Appendix~\ref{sec:apdx_anal_atmos}), but to a much less 
degree.  This issue will be further studied in a separate work.

\subsection*{Error Correlations}

Another issue to address is how any two mixing angles correlate with each 
other.  In principle, the 68$\%$ CL constraints on $\thDA_{ij}$ should be 
expected to be the same as those on $\thA_{ij}$ since they are extracted 
from the same data sample. By virtue of this fact, one can estimate the 
correlation coefficients between any two mixing angles in each representation. 
For two different representations, say $\RepG$ and $\RepH$, any mixing angle 
in $\RepG$, $\thG_{ab}$, can be expressed as a function of the three mixing 
angles, $\thH_{ij}$, in $\RepH$ as:
\begin{equation*}
\thG_{ab} = F(\thH_{12}, \thH_{23}, \thH_{13})\, .
\end{equation*}
By applying the error propagation 
\begin{eqnarray}
&  & \hskip-1.0cm
(\Delta \thG_{ab} )^2  =   \sum_{i \neq j}  
    ( \frac{\partial F}{\partial \thH_{ij}} )^2 (\Delta \thH_{ij} )^2
                                                         \nonumber \\ 
&  &  \hskip-0.6cm
  + 2 \sum_{i \neq j, j \neq k} 
      ( \frac{\partial F}{\partial \thH_{ij}} )
      ( \frac{\partial F}{\partial \thH_{jk}} )
      (\Delta \thH_{ij} ) (\Delta \thH_{jk} ) \rhoH(\thH_{ij}, \thH_{jk})\, ,
\label{eq:AD_corr_coef}
\end{eqnarray}
the correlation coefficients $\rho$ can be analytically solved 
using Eq.~(\ref{eq:AD_corr_coef}). With the best-fit results from the global 
minimum in $\RepD$ and the corresponding translated results in $\RepA$,
along with the average of upper and lower bounds of uncertainties from
$\thD_{ij}$ and $\thA_{ij}$,
one can estimate the correlation coefficients $\rho$ of any two mixing 
angles in representations $\RepA$ and $\RepD$, respectively. 
Two situations are compared, {\it (i)} the 68$\%$ CL constraints taken at 
$\Delta \chi^2 = 2.3$ for all $\thA_{ij}$ (or $\thDA_{ij}$) and $\thD_{ij}$;
and {\it (ii)} the 68$\%$ CL constraints taken at $\Delta \chi^2 = 1.0$ 
for $\thA_{13}$ while keeping the rest at $\Delta \chi^2 = 2.3$.
The results of the estimated correlation coefficients are presented in
Table~\ref{table:corr_coef_AD}.
\begin{table}[th]
\begin{tabular}{cccc}  \hline\hline
Rep.  &  $\rho(\theta_{12}, \theta_{23})$  &  $\rho(\theta_{23}, \theta_{13})$
      &  \vspace{0.1cm} $\rho(\theta_{12}, \theta_{13})$        \\  \hline
\multicolumn{4}{c}
    { Case (i): $\Delta \chi^2 = 2.3$  }                        \\
\multicolumn{4}{c}
    {\underline{
         for all $\Delta \thA_{ij}$ and $\Delta \thD_{ij}$ } }      \\
$\RepA$ &  $ 0.288$  &  $ 0.086$  &  $0.290$                    \\
$\RepD$ &  $-0.854$  &  $-0.114$  & \vspace{0.1cm} $0.623$      \\  \hline
\multicolumn{4}{c}
    { Case (ii): $\Delta \chi^2 = 1.0$ for $\Delta \thA_{13}$ }  \\
\multicolumn{4}{c}
  {\underline{
       but $\Delta \chi^2 = 2.3$ for the rest
             } }                                      \\
$\RepA$ &  $1.861$   &  $-0.554$  &  $3.000$          \\
$\RepD$ &  $-1.082$  &  $-0.409$  &  $0.632$          \\  \hline \hline
\end{tabular}
\caption{The estimated correlation coefficients of any two mixing angles
         in representations $\RepA$ and $\RepD$, where the 68$\%$ CL
         constraints on $\thA_{13}$ and $\thD_{13}$ are assumed by
         $\Delta \chi^2$ in two situations.}
\label{table:corr_coef_AD}
\end{table}

It is interesting to note that the correlations between any two mixing 
angles in representations $\RepA$ and $\RepD$ are bigger if case {\it (ii)} 
is applied. The outcome of the correlations for case {\it (ii)} seems to 
conflict with existing neutrino data which indicate the three mixing angles 
in representation $\RepA$ are nearly decoupled.  However, the results of the 
correlations in case {\it (i)} are more consistent with the nearly-decoupled 
feature among the three mixing angles in representation $\RepA$.  
This suggests that the 68$\%$ CL constraints on $\thA_{13}$ seem to require 
$\Delta \chi^2 = 2.3$ rather than $\Delta \chi^2 = 1.0$ since $\thA_{13}$ 
are also correlated with the other mixing angles 
(cf. Table~\ref{table:corr_coef_AD}).   
In addition, the correlations among the three mixing angles in representation 
$\RepA$ are not completely zero though small, indicating that the three
mixing angles in representation $\RepA$ are not really decoupled.
Nevertheless, the correlations among the three mixing angles in 
representation $\RepA$ are found to be smaller than those in $\RepD$, 
as expected.  
With the estimated $\rhoD$, the 68$\%$ CL constraints on $\thDA_{ij}$ 
for the second minimum are presented in Table~\ref{table:best_fits_DA}.


%
\section{Summary and Outlook}\label{sec:summary}

We have performed global neutrino oscillation data analyses in two 
representations for the neutrino mixing matrix.  Individual data samples 
we used include those from solar, KamLAND, long-baseline accelerator, 
CHOOZ and atmospheric experiments.  We found that individually they do 
not constrain the three mixing angles in representation $\RepD$ as well
as those in representation $\RepA$.
However, the 68$\%$ CL constraints on $\thD_{ij}$ retrieved from combined 
global neutrino data in representation $\RepD$ are almost as good as those 
on $\thA_{ij}$ determined in representation $\RepA$, with the exception of 
$\thD_{12}$ which is a little bit worse than $\thA_{12}$ by 
$\sim$1.5$^{\circ}$.
Nevertheless, the best-fit $\thD_{ij}$ results, as expected, turn out to 
be significantly large. This results provide a higher sensitivity to the 
CP-violating phase determination as can be  seen from the Jarlskog invariant 
quantity.
In addition, the translated best-fit result of each $\thDA_{ij}$ from 
representation $\RepD$ to $\RepA$ are found to be in very good agreement 
with the corresponding angle, $\thA_{ij}$, directly retrieved from 
representation $\RepA$. In other words, $\thDA_{12}$, 
$\thDA_{23}$, and $|\thDA_{13}|$ are respectively consistent with $\thA_{12}$ 
determined from the combined result of solar$+$KamLAND, $\thA_{23}$ obtained 
from the combined result of the {\mbox accelerator$+$CHOOZ$+$atmospheric} data 
sets, and $|\thA_{13}|$ extracted from the combined result of global neutrino 
data in representation $\RepA$. 

As we enter the era of precision neutrino oscillation experiments, 
observables with terms linear in $\sin\theta_{13}$ may be no longer 
negligible in fitting the mixing angles.
The sign of $\sin\theta_{13}$ plays a key role in the determination of 
the sign of $\CPV$, and both $\sin\theta_{13}$ and $\CPV$ are important
in the establishment of the mixing matrix.
In this work we have shown that one has a chance to identify 
the sign of $\thDA_{13}$ through an oscillation parameter fitting performed 
in representation $\RepD$.
We found two local minima when analyzing global neutrino data in 
representation $\RepD$, one corresponding to $\thDA_{13} < 0$ (the global 
minimum) and the other one $\thDA_{13} > 0$ (the second minimum).
A weak preference for negative $\sin \theta_{13}$ is found.
It is interesting to note that the global solar data point to 
$\thDA_{13} > 0$ while all the other data $\thDA_{13} < 0$. 
We do not have interpretation to this difference yet.  One may investigate 
whether the matter (MSW) effects have impacts on the sign of $\thDA_{13}$ 
during the propagation of solar neutrinos through the Sun and the Earth.
This issue will be further studied in a separate work.  
In addition, we also provide the information for the correlation of any 
two mixing angles. It is found that the correlations in representation 
$\RepA$ are less than those in $\RepD$, but the three mixing angles 
are not completely decoupled in representation $\RepA$.   

In conclusion,
owing to the strong correlations among the three mixing angles
in the new parametrization, the advantages of doing the global neutrino
oscillation analysis using data from past, current, and near future
neutrino oscillation experiments shall become manifest.


%
\section{Acknowledgements}

First of all, we are indebted to Jen-Chieh Peng for inspiring this idea and
his valuable discussion and suggestions. MH and SDR are grateful to 
Bruce~T.~Cleveland, Bernhard~G.~Nickel, Jimmy Law, and Ian~T.~Lawson for 
their useful discussion and suggestions, especially to Bernhard~G.~Nickel 
and Jimmy Law for their generosity in providing us with the adiabatic codes. 
We thank Hai-Yang Cheng, Guey-Lin Lin, and Jen-Chieh Peng for careful reading 
the manuscript and valuable comments.  MH and WCT are thankful to Pisin Chen 
for his support and useful comments. SDR appreciates Pisin Chen for his 
hospitality and financial support during her stay at NTU, Taiwan. Last 
but not least, WCT and HT thank Yung-Shun Yeh for his help with installing 
and running the NUANCE package. 
This research was supported by Taiwan National Science Council under Project
No. NSC 98-2811-M-002-501 and NSC 99-2811-M-002-064, Canadian Natural Sciences
and Engineering Research Council, and Fermi Research Alliance, LLC under the 
U.S. Department of Energy contract No. DE-AC02-07CH11359. 
This work was also made possible by the facilities of the Shared Hierarchical 
Academic Research Computing Network (SHARCNET) and \mbox{Compute/Calculation} 
Canada and by the FermiGrid facilities for the availability of their resources. 


%
\appendix
\section{Solar Neutrino Sector}\label{sec:apdx_anal_solar}
%

\subsection{Cl and Ga Rate Neutrino Data}

The rates for the Chlorine and Gallium experiments are expressed in terms
of SNUs (1 Solar Neutrino Unit = one interaction per 1036 target atoms per
second).  The predicted rate for the Chlorine/Gallium experiment is given by
\begin{equation}
R_{Cl/Ga} = \int^{\infty}_{E_{th}} dE_{\nu}
         \phi_{\nu_e}(E_{\nu}) P_{ee}(E_{\nu}) \sigma_{Cl/Ga}(E_{\nu})  ,
\label{eq:solar_ClGa_1}
\end{equation}
where $E_{\nu}$ and $E_{th}$ are respectively the neutrino energy and
the threshold energy for neutrinos captured by chlorine or gallium,
$\phi_{\nu_e}(E_{\nu})$ is the flux of electron neutrinos arriving at
the detector, including the flux from all solar neutrino reactions,
$P_{ee}(E_{\nu})$ is the survival probability of $\nu_e \rightarrow \nu_e$,
$\sigma(E_{\nu})$ is the cross section of electron neutrino with the
target (either chlorine or gallium).
The Cl rate of 
$2.56 \pm 0.l6 {\rm(statistical)} \pm 0.16 {\rm(systematic)}$ SNU 
from Ref.~\cite{Cleveland:1998nv} and the Ga rate of $66.1 \pm 3.1$ SNU from 
SAGE, Gallex, and GNO \cite{Abdurashitov:2009tn} are used in this analysis.

\subsection{Borexino Rate Neutrino Data}

The predicted rate calculation of $^{7}$Be solar neutrinos measured in 
Borexino experiment is calculated as 
\begin{equation}
R_{Bx} = R^{0}_{Bx} 
         \left\{ S_{e}(E_{\nu}, E_e) p_e + 
                 S_{x}(E_{\nu}, E_e) (1 - p_e) \right\} ,
\label{eq:solar_Bx_1}
\end{equation}
where $R^{0}_{Bx}$ is the expected rate for non-oscillated solar $\nu_e$
that is $74 \pm 4$ counts/(day $\times$ 100 ton), $S_{e,x}(E_{\nu},E_e)$ 
describe the probability that the elastic scattering of a $\nu_{e,x}$
of energy $E_{\nu}$ with electrons produces a recoil electron of energy 
$E_e$, and $p_e$ is the survival probability of $\nu_e \rightarrow \nu_e$.
The measured rate of 
$49 \pm 3_{stat} \pm 4_{syst}$ counts/(day $\cdot$ 100 ton) from 
Borexino experiment \cite{Arpesella:2008mt} is adopted in this analysis.

\subsection{Super-Kamiokande Solar Neutrino Data}

The measured day/night spectra of Super-Kamiokande (SK) phases I \& II 
\cite{Hosaka:2005um,Cravens:2008zn} are employed in this analysis. 
To fit for the oscillation parameters, we follow the methods described 
in Ref.~\cite{Hosaka:2005um} to build up the predicted rate calculation 
for SK solar day/night spectra, which is given by 
\begin{eqnarray}
R_{SK} &=& \int^{E_1}_{E_0} dE  \int_{E_{\nu}}  dE_{\nu} I(E_{\nu}) 
           \int_{E_e} dE_{e} R(E_e, E) \times              \nonumber \\
       & & \left\{ S_{e}(E_{\nu}, E_e) p_e + 
                   S_{x}(E_{\nu}, E_e) (1 - p_e) \right\}  ,
\label{eq:solar_SK_1}
\end{eqnarray} 
where $I(E_{\nu})$ is the spectrum of $^{8}$B or {\it hep} neutrino,
$R(E_e, E)$ is the detector response function representing the probability
that a recoil electron of energy $E_e$ is reconstructed with energy $E$, 
and $S_{e,x}(E_{\nu}, E_e)$ and $p_e$ have the same meanings as given in 
previous paragraph. Both rates due to $^{8}$B and {\it hep} interactions 
are taken into consideration in this work.

\subsection{SNO Solar Neutrino Data}

For SNO solar data, the fraction of the extracted $CC$ (charged current) 
spectra as a function of one unoscillated SSM \cite{Bahcall:2004pz} 
(Fig.~28 in Ref.~\cite{Aharmim:2009gd}) is used for the mixing parameter 
fitting. 
The theoretical calculation of the $CC$ flux fraction for a set of oscillation 
parameter in order for comparison with SNO solar data is described as follows. 
The $CC$ flux fraction in the electron kinetic energy between $T_1$ and $T_2$,
denoted as $f(T_1, T_2)$, is calculated as the number of events observed in 
the data set contributed to $CC$ interactions by the signal extraction with 
electron kinetic energies between $T_1$ and $T_2$ divided by the number of 
all $CC$ events that would be observed above the threshold kinetic energy, 
$T_{th}$. It is formulated as~\cite{Aharmim:2005gt}  
\begin{eqnarray}
& & \hskip-0.45cm 
  f(T_1, T_2) \equiv       \nonumber \\
& & \hskip-0.45cm
  \frac{\int^{\infty}_0  \int^{\infty}_0  \int^{T_{2}}_{T_{1}}
         \phi(E_{\nu}) P_{ee}(E_{\nu}) \frac{d \sigma}{dT_e}(E_{\nu}, T_e)
         R(T_e, T^{'}_{e}) dE_{\nu} dT_e dT^{'}_{e}
       }
       {\int^{\infty}_0 \int^{\infty}_0 \int^{\infty}_{T_{th}}
         \phi(E_{\nu}) \frac{d \sigma}{dT_e}(E_{\nu}, T_e)
          R(T_e, T^{'}_{e}) dE_{\nu} dT_e dT^{'}_{e}
       }                   \nonumber \\ 
\label{eq:solar_SNO_1}
\end{eqnarray}
where $P_{ee}$ is the survival probability of $\nu_e \rightarrow \nu_e$, 
$\phi(E_{\nu})$ includes both fluxes of $^{8}$B and {\it hep} solar neutrinos
as a function of neutrino energy, $d\sigma / dT_e$ is the differential cross 
section for $CC$ interactions, and $R(T_e, T^{'}_{e})$ is the energy resolution 
function, describing the probability of seeing an apparent kinetic energy 
$T^{'}_{e}$ for a given true energy $T_e$, whose expression can be found 
in Ref.~\cite{Aharmim:2009gd}.  

\subsection{Global Solar Neutrino Data}

To fit for oscillation parameters using global solar data, the combined 
$\chi^2$ for the observables and predictions is given by
\begin{eqnarray}
& & \hskip-1.0cm
  \chi^2_{solar} =   \nonumber \\
& & \hskip-0.8cm 
   \sum^{N}_{i, j = 1} ( \mathcal{O}_{i} - \mathcal{O}^{exp}_{i} )
                           [ \sigma^2_{i j}(tot) ]^{-1} 
                           ( \mathcal{O}_{j} - \mathcal{O}^{exp}_{j} ) ,
\label{eq:apdx_solar_chi2}
\end{eqnarray}
where $i$, $j$ are indices to sum over all the observables.
The quantities $\mathcal{O}_{i}$ 
\begin{equation*}
\mathcal{O}_{i} = \mathcal{O}_{i}(\Delta m^2, \tan^2 \theta, \ldots)
\end{equation*}
is the theoretical prediction for that observable, and 
$\mathcal{O}^{exp}_{j}$ represents a series of observables from a number 
of solar neutrino experiments that include rate measurements from Cl 
and Ga (e.g. Homestake, Gallex, GNO, and SAGE) experiments, rate of 
$^{7}$Be solar neutrino measurement from Borexino experiment, a number 
of spectral shape measurements from SK and SNO as mentioned above.
The total error matrix, $\sigma^2(tot)$, is a sum of contributions from 
the rate and spectral measurements, which is given as
\begin{eqnarray}
& & \hskip-1.0cm
    \sigma^2(tot) =  \nonumber \\
& & \hskip-0.8cm
    \sigma^2(exp) + \sigma^2_R + \sigma^2_R(Bx)
                 + \sigma^2_{S}(SNO) + \sigma^2_{S}(SK) ,
\label{eq:solar_sigma_2}
\end{eqnarray} 
where $\sigma^2(exp)$ is a diagonal matrix containing the statistical 
and systematic errors from the rate measurements and the statistical 
errors from the spectral measurements;  
$\sigma^2_R$ is the rate error matrix, handling the correlations 
between the rate CL and Ga experiments using the procedure described 
in Ref.~\cite{Fogli:1994nn};  
$\sigma^2_R(Bx)$ is treated as independent from $\sigma^2_R$ in our 
analysis. Furthermore, the spectral correlation matrix $\sigma^2_{S}(SNO)$
for SNO measurements is assumed to be uncorrelated with $\sigma^2_{S}(SK)$
for SK measurements, which are uncorrelated with the CL, Ga, and 
Borexino measurements. One can refer to 
Refs.~\cite{Fogli:1994nn, Fogli:1999zg, Garzelli:2000tn, 
Garzelli:2001zu} for detailed discussion on the covariance error matrix,
$\sigma^2(tot)^{-1}$.
By minimizing $\chi^2_{solar}$ with three-flavor neutrino oscillation scheme 
while fixing $\Dmbc = 2.4 \times10^{-3}$ eV$^2$ and $\sinthbc = 0.5$, the
values of $\Dmab$, $\tanthab$, and $\sinthac$ were found to reproduce the
results of global solar data as reported in Ref.~\cite{Aharmim:2009gd} 
when the conventional mixing matrix parametrization is used.


%
\section{Reactor Neutrino Sector}\label{sec:apdx_anal_reactor}
%

\subsection{KamLAND Reactor $\barnue$ Data}

For KamLAND reactor data, the prompt energy spectrum of $\bar{\nu}_e$
events of energy (as shown in Fig.~(1) of Ref.~\cite{Gando:2010aa}) 
is used. 
We follow the strategy described in Ref.~\cite{Fogli:2005qa} for this 
analysis. Since the distance from the reactor source to the KamLAND 
detector is $\sim$180 Km, the treatment of vacuum oscillation is 
considered in this analysis. With a little modification, the number 
of expected events per unit of the prompt position energy is given by 
%
\begin{eqnarray}
& & \hskip-1.0cm
N(T^{'}_{e}) = \xi \Delta t \cdot \epsilon(T^{'}_{e}) \cdot
             \int dE_{\nu} \frac{d \phi}{d E_{\nu}}  \nonumber \\
& &  \times \int dT_e \frac{ d\sigma(E_{\nu}, T_e) }{ dT_e }
                                    R(T_e, T^{'}_{e}) \; .
\label{eq:reactor_KL_1}
\end{eqnarray}
The values $\xi = 5.98 \times 10^{32}$ protons and $\Delta t = 2135$ days 
are the total number of target protons and livetime, respectively.
$\epsilon(T^{'}_{e})$ is the detection efficiency as a function of 
measured prompt energy, $T^{'}_{e}$ .
$d \phi / d E_{\nu}$ is the time-averaged differential neutrino flux at 
the KamLAND detector, where the fission flux as a function of distance 
can be found at the website \cite{KL_2nd_fission_flux} and the relative 
fission yields, $^{235}$U:$^{238}$U:$^{239}$Pu:$^{241}$Pu, is taken from 
\cite{Gando:2010aa}, and flux from Korean reactors is not taken into account
since only $\sim$3\% of the total flux contributes to KamLAND signal.
In the presence of oscillation, each $j$th reactor term in 
$d \phi / d E_{\nu}$ must be multiplied by the corresponding neutrino 
survival probability $P_{ee}(E_{\nu}, L_j)$ for neutrinos of energy 
$E_{\nu}$ and taveling distance of $L_j$ between each reactor source 
and the KamLAND. 
$R(T_e, T^{'}_{e})$ is the energy resolution function with Gaussian width 
equal to 6.4\%$\sqrt{T_e/MeV}$, which has the same meaning as described
in the solar sector.
$ d\sigma(E_{\nu}, T_e) / dT_e $ is the inverse beta decay cross section 
\cite{Fogli:2005qa, Vogel:1999zy}. 

To fit for the oscillation parameters, the $\chitwo$ function is of the
Gaussian form \cite{Fogli:2005qa} that includes both the total number 
of events and the spectral shape: 
\begin{eqnarray}   
\chi^2_{KL} &=&            
  \left( \frac{N^{theo}_{tot} - N^{obs}_{tot}}{\sigma^{rate}_{tot}} \right)^2  
                                                \nonumber \\
& & 
  + \sum_i 
   \left( \frac{N^{theo}_i - N^{obs}_i}{\sigma^{rate}_i} \right)^2 
  + \left( \frac{\alpha}{\sigma^{\alpha}} \right)^2 ,
\label{eqn:chi_KL}
\end{eqnarray}
where $N^{obs}_{tot} = 1780$ is the total number of observed events after 
substracting all background events, $N^{theo}_{tot}$ is the predicted total
events; $N^{theo}_i$ and $N^{obs}_i$ are the numbers of observed events and 
predicted events respectively in $i$th energy bin, the total error, 
$\sigma^{rate}_{tot}$, is the quadrature sum of the statistical and 
systematic uncertainties, and $\sigma^{rate}_i$ is the error for $i$th 
bin by adding statistical and systematic uncertainties in quadrature.
Minimizing the $\chi^2_{KL}$ while fixing $\Dmbc = 2.4\times10^{-3}$ eV$^2$ 
and $\sinthbc = 0.5$, the values for $\Dmab$, $\tanthab$, and $\sinthac$ 
were found to reproduce the KamLAND results in Ref.~\cite{Gando:2010aa}
if the conventional mixing matrix parametrization is applied.

\subsection{CHOOZ Reactor $\barnue$ Data}

The CHOOZ results~\cite{Apollonio:2002gd} reported the observed positron 
energy from the neutrino interactions. For this experiment, the distance 
between the detector and the neutrino source is relatively small (1 km) 
and thus the neutrino propagation through vacuum can be used to calculate 
the anticipated positron spectrum. 
Using the procedures reported in Ref. \cite{Apollonio:2002gd}, the expected 
positron yield for the $k$-th reactor and the $j$-th energy spectrum 
bin is parametrized as 
\begin{eqnarray}
& & \hskip-0.8cm
\bar{X}(E_j, L_k, \theta, \delta m^2) =
   \tilde{X}(E_j) \cdot \bar{P}(E_j, L_k, \theta, \delta m^2_{32}),  
                                                            \nonumber \\
& & \hskip2.3cm 
   (j = 1, \ldots , 7, \hskip0.2cm  k = 1, 2) ,
\label{eq:reactor_Chooz_1}
\end{eqnarray}
where $\tilde{X}(E_j)$ is the distance-independent positron yield for 
no presence of neutrino oscillation, $L_k$ is the reactor-detector 
distance, and $\bar{P}(E_j, L_k, \theta, \delta m^2_{32})$ is the 
oscillation probability averaged over the energy bin and the core 
sizes of the detector and the reactor. For the experimental data, 
the results reported in Table 8 in Ref.~\cite{Apollonio:2002gd} were 
used.  These results consist of seven positron energy bins for each 
of the two reactors for a total of 14 bins.
The covariant matrix $V^{-1}_{ij}$ defined in equation 54 in
Ref.~\cite{Apollonio:2002gd} is applied to the $\chi^2$ function. 
This matrix accounts for any correlations between the energy bins 
of the two reactors.  We then minimize the $\chi^2$ function with 
respect to the neutrino oscillation parametrization and we can reproduce 
the CHOOZ results as presented in Ref.~\cite{Apollonio:2002gd} when 
the conventional mixing matrix parametization is applied.


%
\section{Long-Baseline Accelerator Neutrino Sector}\label{sec:apdx_anal_accel}
%

\subsection{MINOS}\label{apdx_accel_MINOS}

The analysis for MINOS used results from the $\numu$ disappearance
channel~\cite{Adamson:2011ig} and the $\nue$ appearance 
channel~\cite{MINOS_nue_app_2011}. Results from both channels consisted 
of the reconstructed neutrino energy as seen in the far detector.
The distance of 735 km between the near and far detectors was considered 
to be too small for matter effects to have any significant contribution 
to the results. Vacuum oscillation is thus applied in this analysis. 

The predicted number of neutrinos at the far detector is calculated as:
\begin{equation}
  N^{osc} = f( P_{\mu\mu} + \epsilon_{\tau} P_{\mu\tau} ) N^{no\ osc} 
            + g N^{NC}
\label{eqn:chi_minos_k2k_1}
\end{equation}
where $P_{\mu\mu}$ is the survival probability for $\numu$, 
$P_{\mu\tau}$ is the probability for $\nutau$ appearance, $\epsilon_{\tau}$ 
is the $\nutau$ detection efficiency, $N^{NC}$ is the $NC$ (neutral current) 
spectrum, $f$ is the normalization factor, and $g$ is the $NC$ scaling factor.
Like the $\numu$ data, the $\nue$ data consists of events binned by 
reconstructed energy over a range of 1 to 8 GeV. The predicted neutrino 
spectrum of $\nue$ appearance channel is calculated using:
\begin{equation}
  N^{osc} = f^{'} P_{\mu e} N^{no\ osc} + g^{'} N^{NC}
\label{eqn:chi_minos_k2k_2}
\end{equation}
where $P_{\mu e}$ is the probability for $\nue$ appearance, and $f^{'}$ 
$(g^{'})$ has the same meaning as $f$ $(g)$ but with a different value.  
The value for $N^{no\ osc}$ is taken from~\cite{Adamson:2011ig}.

The $\chitwo$ function is of the Poisson form and includes both the spectral
shape and the total number of neutrino events, which is given by:
\begin{eqnarray}
\chi^2 &=& 
    2[ N^{osc}_{tot} - N^{obs}_{tot} ( 1+\ln( N^{osc}_{tot}/N^{obs}_{tot} )) ]  
                                                        \nonumber \\
  & & + \sum_{i} 2[ N^{osc}_{i} - N^{obs}_{i} ( 1+\ln( N^{osc}_{i}/D_i )) ]    
                                                        \nonumber \\
  & & + \sum_{j} \left( \frac{\xi_j}{\sigma_j} \right)^2 ,
\label{eqn:chi_minos_k2k}
\end{eqnarray}
where $N^{osc}_{tot}$ and $N^{obs}_{tot}$ are the predicted and detected 
total number of neutrinos respectively;  $N^{osc}_{i}$ and $N^{obs}_{i}$  
are the predicted and detected number of neutrinos in the $i^{th}$ energy 
bin.  The systematics for the energy scale, normalization factor and scale 
factor for the $NC$ events are represented by $\xi_j$ with uncertainty 
of $\sigma_j$.
Minimizing the $\chitwo$ for the $\numu$ disappearance channel while
fixing $\Dmab = 7.67\times10^{-5}$ eV$^2$, $\tanthab = 0.427$, 
$\sinthac = 0.0$ and assuming a normal mass hierarchy, the values for 
$\Dmbc$ and $\sinthbc$ where found to reproduce the MINOS results 
in Ref.~\cite{Adamson:2011ig} when the mixing matrix parametrization
$\RepA$ is applied.

\subsection{K2K}\label{apdx_accel_K2K}

The results from K2K~\cite{Ahn:2006zza} used in this analysis were the 
reconstructed $\numu$ energy spectrum for one-ring $\mu$-like sample.
Like MINOS, the K2K's 250-km oscillation length is considered to be 
too short to manifest matter enhanced oscillation effects. Vacuum 
oscillation can thus be applied to this analysis.

The predicted no-oscillation neutrino spectrum represents the true 
neutrino energy spectrum at the near detector. 
To extract the predicted far detector oscillation spectrum from the 
no-oscillation spectrum, the neutrino interaction cross sections and
detection efficiencies must be applied.  Furthermore, both $CC$ and 
$NC$ events are present in the neutrino spectrum and hence are accounted 
for separately. The expected number of neutrino events for oscillation 
neutrinos is given by
\begin{equation}
N^{osc}_j = N^{no\ osc}_j 
   \left[ \epsilon_{cc} P_{cc}\sigma_{cc} + \epsilon_{nc}\sigma_{nc} \right]
\end{equation}
where $\sigma_{cc}$ and $\sigma_{nc}$ are the cross sections for the 
$CC$ and $NC$ interactions respectively, $\epsilon_{cc}$ and $\epsilon_{nc}$ 
are the detection efficiencies for the $CC$ and $NC$ interactions, and 
$P_{cc}$ is the survival probability for $\nu_{\mu}$. 
The energy response function is applied to obtain the probability of seeing 
the measured energy for a given true energy, same meaning as employed in the
solar analyses.  The form of the energy response function is employed from 
Ref.~\cite{Fogli:2003th}:
\begin{equation}
N^{obs} = \sum_{i} N^{osc}_{i} * 
            e^{ -\frac{1}{2} ( (E_{obs} - E_{i} + 0.05)/\sigma )^2 }
\end{equation}
where $E_{obs}$ and $E_{i}$ are the measured and true neutrino energies
respectively, and $\sigma$ is defined as \cite{Ahn:2006zza}:
\begin{equation*}
\sigma = 0.2 E_{i} ( 1-e^{ (0.2-E_{i}) / 0.8} )
\end{equation*}

The K2K analysis uses the same $\chitwo$ function as that used in the MINOS 
analysis.  By minimizing the $\chitwo$ for the $\numu$ while fixing 
$\Dmab = 7.67 \times10^{-5}$ eV$^2$, $\tanthab = 0.427$, 
$\sinthac = 0.0$, and assuming a normal mass hierarchy, the values for 
$\Dmbc$ and $\sinthbc$ were found to reproduce the K2K results in
Ref.~\cite{Ahn:2006zza} when the conventional mixing parametrization 
is applied. 

%
\section{Atmospheric Neutrino Sector} \label{sec:apdx_anal_atmos}
%

\subsection{Super-Kamiokande Atmospheric Data}

The Super-Kamiokande (SK) detector is located deep under the peak of Mt 
Ikenoyama, where the 1,200-m rock overburden can reduce the flux of cosmic 
rays reaching the detector down to about 3 Hz (see e.g. 
Ref.~\cite{Wendell:2008zz}). Atmospheric neutrinos penetrating the Earth 
interact with the nucleus or nucleons in the SK water tank or in the 
surrounding rock, giving rise to partially contained (PC), fully contained 
(FC), or upward-going muon (UP$\mu$) events. Two and three-flavour neutrino 
oscillation analyses~\cite{Wendell:2010md,Hosaka:2006zd,Ashie:2005ik}
have been performed using the accumulated data from SK-I, II and III phases.

We try to reproduce the results of the 3-flavour analysis published in 
Ref.~\cite{Hosaka:2006zd} to include in our global analysis. Therein, the 
1489 live-days of PC and FC, and the 1646 days UP$\mu$ data collected during 
the SK-I period (1996-2001) are used. Based on the types of the out-going 
leptons, their energy deposited in the SK detector, and their zenith angle 
($-1 < \cos \theta_{\rm zenith} < 1$ for PC and FC; 
 $-1 < \cos \theta_{\rm zenith} < 0$ for UP$\mu$),
the selected events are divided in 370 bins. In our work, however, lack of 
information forced us to follow the approach of Ref.~\cite{Fogli:2003th}, 
where only 55 energy and zenith angle bins of SK atmospheric data are used.

Here we briefly describe our analysis.
First, the NUANCE package~\cite{Casper:2002sd} is used to simulate atmospheric
neutrino events assuming no oscillation effects. For achieving good statistics 
we have run the simulations for 200-year operation time. Neutrino oscillation 
effects in the atmosphere and inside the Earth are then incorporated by the 
"weighting" factors~\cite{Wendell:2008zz} as:
\begin{eqnarray}
& & \hskip-1.0cm
w_{e} = P(\nu_{e} \rightarrow \nu_{e}) + 
          \frac{\phi_{\mu}}{\phi_{e}}P(\nu_{\mu} \rightarrow \nu_{e})\, ,
          \hspace{0.4cm} {\rm for} \hspace{0.1cm} \nu_{e}\, , \nonumber \\
& & \hskip-1.0cm
w_{\mu} = P(\nu_{\mu} \rightarrow \nu_{\mu}) + 
            \frac{\phi_{e}}{\phi_{\mu}}P(\nu_{e} \rightarrow \nu_{\mu})\, ,
            \hspace{0.4cm} {\rm for} \hspace{0.1cm} \nu_{\mu}\, . 
\label{eq:weighting_factors}
\end{eqnarray} 
For the (energy and angular-dependent) incident atmospheric neutrino 
fluxes $\phi_{e, \mu}$, we also adopt the Honda three-dimensional 
calculation~\cite{Honda:2004yz}. We take into account the effects 
when neutrinos oscillating in matter following the prescription of 
Ref.~\cite{Barger:1980tf}, and approximate the Earth as a four-layer 
division as Ref.~\cite{Hosaka:2006zd} did. Each layer has a constant 
density
(inner core: $R \leq 1221$ km, $\rho=13.0$ g/cm$^{3}$;
 outer core: $1221 < R \leq 3480$ km, $\rho=11.3$ g/cm$^{3}$;
 mantle: $3480 < R \leq 5701$ km, $\rho=5.0$ g/cm$^{3}$;
 crust: $5701 < R \leq 6371$ km, $\rho=3.3$ g/cm$^{3}$).
We apply similar criteria and cuts on the kinematics of the simulated 
events so that we achieve the same selection efficiency as 
Ref.~\cite{Hosaka:2006zd} does. 

Furthermore, one can include the systematics by using the "pull 
technique"~\cite{Fogli:2003th,Fogli:2002pt}.
Due to correlated systematic uncertainties, the event rate in the $n$-th 
bin predicted by the MC simulation, $R^{MC}_n$, is shifted by an amount as
\begin{equation*}
  R^{MC}_n \rightarrow \widetilde{R_{n}^{MC}} \equiv R_{n}^{MC} 
           \left( 1 + \sum_{k=1}^{11} S_{n}^{k}\xi_{k} \right)\, .
\end{equation*}
Here $S_{n}^{k}$ is the 1$\sigma$ error associated to the $k$-th source 
of systematics, and $\xi_{k}$'s are a set of univariate Gaussian random 
variables.  Ref.~\cite{Fogli:2003th} gives a summary and detailed discussion 
of the 11 systematics sources, which we also use in our work.
The $\chi^2$ function thus becomes
\begin{eqnarray}
& & \hskip-1.0cm
\chi_{\rm SK}^{2} = {\min}_{\{\xi_k\}} 
  \left[ \sum_{n=1}^{55}\left( \frac{\widetilde{R_{n}^{MC}}-R_{n}^{ex}}
                    {\sigma_{n}^{stat}} \right)^{2}  \right.   \nonumber \\
& & \hskip1.1cm
    \left.  + \sum_{k,h=1}^{11}\xi_{k}[\rho^{-1}]_{kh}\xi_{h} \right]\, , 
\label{eq:ChiSq_formula}  
\end{eqnarray}
for the observed event rate $R_{n}^{ex}$ in each bin.
Here $\sigma_{n}^{stat}$ is the statistical error of the $n$-th bin, and
$\rho^{-1}$ the inverse of the correlation matrix, the value of which can 
also be found in Ref.~\cite{Fogli:2003th}.

We minimize the $\chi^{2}_{\rm SK}$ value with respect to the oscillation 
parameters by solving $\frac{\partial \chi^{2}}{\partial \xi_{i}}=0$, which 
is equivalent to solving a set of linear equations:
\begin{eqnarray}
& & \hskip-0.5cm
 \sum_{k=1}^{11} \left[ \sum_{n=1}^{55}\left(\frac{R_{n}^{MC}}
                       {\sigma_{n}^{stat}}\right)^{2}\times 
                S_{n}^{i}S_{n}^{k}+ \rho^{-1}_{ik}\right]\xi_{k}  \nonumber \\
& & \hskip0.5cm
 = \sum_{n=1}^{55}\left[ \frac{R_{n}^{ex}R_{n}^{MC}-(R_{n}^{MC})^{2}}
                        {(\sigma_{n}^{stat})^{2}} \right]S_{n}^{i}\, .
\label{eq:LiEq_to_minize_Xi}
\end{eqnarray}

Under normal hierarchy and $\RepA$ parametrization, our best-fit values for 
($\Delta m^{2}_{32}$, $\sin^{2}\theta_{23}$, $\sin^{2}\theta_{13}$) are 
($2.4\times10^{-3}\mbox{ eV}^{2}$, $0.45$, $0.0$), and
the $90\%$ confidence levels ($\Delta \chi^{2}=4.61$) are 
$1.5\times10^{-3}\mbox{ eV}^{2}<\Delta m^{2}_{32}<3.3\times10^{-3}
\mbox{ eV}^{2}$, 
$0.35<\sin^{2}\theta_{23}<0.59$, and $\sin^{2}\theta_{13}<0.17$, 
which is consistent with those published in Ref.~\cite{Hosaka:2006zd}. 
%

\subsection{SNO Atmospheric Data}

Thanks to its deep location and flat overburden, the SNO detector can 
observe atmospheric neutrinos over a wide range of zenith angle, via 
their charged-current interactions in the surrounding rock.
Angular distribution of through-going muons having 
$-1 \leq \cos \theta_{\rm zenith} < 0.4$
can be used to infer neutrino oscillation parameters and incident neutrino 
flux, while data above this cutoff provide access to the study of the 
cosmic-ray muon flux.

Below we describe briefly our approach to including the latest SNO 
atmospheric neutrino data from Ref.~\cite{Aharmim:2009zm}.
Details of the analyses can be found in T.~Sonley's PhD 
thesis~\cite{Sonley:2009}.  
We first use the NUANCE package~\cite{Casper:2002sd} to simulate atmospheric 
neutrino induced through-going muons and muons generated in SNO detector's 
D$_2$O and H$_2$O regions for 100 year operation time, assuming no oscillation. 
Oscillation effects are then added with the help of the "weighting" factor.
For the incident atmospheric neutrino fluxes, we adopt the Bartol 
three-dimensional calculation~\cite{Barr:2004br}. Due to lack of a detector 
simulation package such as the SNOMAN, we apply simple cuts on the 
kinematics of simulated events, separately for different event types 
($\nu_\mu$-induced through-going muons, $\nu_\mu$ water interactions, 
and $\nu_\mu$ and $\nu_e$ internal interactions). 
We require (i) the impacat parameter $b \leq 830$~cm;
(ii) the muon energy when entering the detector $E_{\mu} \gtrsim$ 1 GeV, 
and adjust the "trigger efficiency" so that the yearly event rate match that 
given in Ref.~\cite{Aharmim:2009zm}.
The trigger efficiency we defined summarises all other instrumental cuts which 
we are not able to apply. In addition, the systematics are taken into account 
using the generalised "pull technique".  
Following Ref.~\cite{Aharmim:2009zm,Sonley:2009}, the likelihood function used 
in our analysis is 
\begin{eqnarray}
& & \hskip-0.8cm
   L_{\rm total} \equiv - \ln {\cal L} =                    \nonumber \\
& & \hskip-0.1cm
   2 \sum_{{\rm bins}\, i} \left[N^{\rm data}_i\, \ln 
       \frac{N^{\rm data}_i}{N^{\rm MC}_ {0\, , i}}\, 
   + \left(N^{\rm MC}_{0\, ,i} - N^{\rm data}_i \right) \right]\,  \nonumber \\
& & \hskip-0.1cm
   - \vec{\alpha}_{\rm min}\, {\cal S}^2\, \vec{\alpha}_{\rm min}\, ,
\label{eq:apdx_sno_atmos_1}
\end{eqnarray}
where $N^{\rm data}_i$ is the measured event number in zenith angle bin $i$, 
and $N^{\rm MC}_{0\, , i}$ denotes the expected event number in bin $i$ with 
the systematics $\vec{\alpha}$ equal to zero. We include 8 systematics error
sources, and calculate their values which minimise the likelihood function
\begin{equation}
   \vec{\alpha}_{\rm min} = 
     \sum_i\, \vec{\beta}^{\cal T}_i\, \left(N^{\rm data}_i -
            N^{\rm MC}_{0\, ,i} \right)\, {\cal S}^{-2}\, .
\label{eq:apdx_sno_atmos_2}
\end{equation}
Here the coefficients $\vec{\beta}_i$ are available in Ref.~\cite{Sonley:2009},
and the new error matrix is 
\begin{equation}
   {\cal S}^2 = \sigma^{-2} + \sum_i\, N^{\rm data}_i\, \vec{\beta}_i\, \times 
   \vec{\beta}^{\cal T}_i\, 
\label{eq:apdx_sno_atmos_3}
\end{equation}
with $\sigma^{-2}$ the diagonal error matrix whose entries represent the size 
of the systematic errors.

We first compare the results of our two-flavour analysis with those in 
Ref.~\cite{Aharmim:2009zm,Sonley:2009}. Normalisation of the Bartol 
three-dimensional atmospheric neutrino flux $\Phi_0$ is determined 
simultaneously with the neutrino oscillation parameters.
Our best fit values for $(\Dmbc\, , \sin^2 2\theta_{23}\, , \Phi_0)$ are
$(2.0 \times 10^{-3}\, , 1.0\, , 1.30)$ in the normal hierarchy scheme.
These are close to those obtained in Ref.~\cite{Sonley:2009}, while still 
within $1\sigma$ of those from Ref.~\cite{Aharmim:2009zm}.
To save CPU time, we then do the three-flavour analysis by keeping $\Phi_0$ 
fixed at 1.30.  Matter effects induced when neutrinos pass through different 
Earth layers are considered following the prescription of 
Ref.~\cite{Barger:1980tf}.  As expected, SNO atmospheric neutrino data are 
not sensitvie to $\theta_{13}$. Since, unlike the Super-Kamiokande experiment,
SNO only observes muons plus a few possible $\nu_e$-induced internal events.


%
\section{$(\thA_{ij}, \CPVA)$ Solutions in Terms of $(\thD_{ij}, \CPVD)$ }
     \label{apdx:Gfit_NumParamSol_DA}

Following the procedure presented in Ref.~\cite{Huang:2011by}, the 
transformation of $(\thD_{ij}, \CPVD)$ from representation $\RepD$ to $\RepA$
is again briefly summarized here by only listing the solutions of the nine
parameters.

To solve for $(\thA_{ij}, \CPVA)$ in representation $\RepA$ when the 
$(\thD_{ij}, \CPVD)$ are known in representation $\RepD$, begin with 
\begin{equation}
U = RWR(\thA_{ij}, \CPVA) = D^L \cdot RWR(\thD_{ij}, \CPVD) \cdot D^R \; ,
\end{equation}
\label{eq:Append_D_DA_1}
where
\begin{eqnarray}
D^L(\PhiA_{Li}) 
    &=& diag \left( e^{i\PhiA_{L1}}, e^{i\PhiA_{L2}}, e^{i\PhiA_{L3}} \right)  
                                                     \; ,  \nonumber \\
D^R(\PhiA_{Ri})
    &=& diag \left( e^{i\PhiA_{R1}}, e^{i\PhiA_{R2}}, 1  \right)   \; .  
\label{eq:Append_D_DA_2}
\end{eqnarray}
Through the nine real parts and the nine imaginary parts of 
Equation~(\ref{eq:Append_D_DA_1}), the solutions to the nine parameters are 
listed as follows:
\begin{eqnarray}
\PhiA_{L2} &=& 0 ,  \nonumber \\
\PhiA_{L1} + \PhiA_{R1} &=& 0 . 
\label{eq:Append_D_DA_41}
\end{eqnarray}
\begin{equation}
\sin^2{\PhiA_{L3}} = \frac{ (a~\sin\CPVD )^2}
                          { (a~\sin\CPVD )^2 + (a~\cos\CPVD - b)^2}  \; ,
\label{eq:Append_D_DA_42}
\end{equation}
where $a = \sD_{23} \sD_{12} \sD_{13}$ and $b = \cD_{23} \cD_{13}$ .
\begin{equation}
\sin^2{\PhiA_{15}} = 
      \frac{ (a^{'} \sin\CPVD )^2 }
           { (a^{'} \sin\CPVD )^2 + (a^{'} \cos\CPVD - b^{'})^2 } \; ,
\label{eq:Append_D_DA_43}
\end{equation}
where $a^{'} = \cD_{23} \sD_{12} \cD_{13}$,  $b^{'} = \sD_{23} \sD_{13}$,
and $\PhiA_{15} \equiv \PhiA_{L1} + \PhiA_{R2}$. 
With these phases, the three mixing angles can be extracted in representation 
$\RepA$:
\begin{eqnarray}
& & \hskip-1.2cm
  \tan \thA_{23} =
       \frac{ \sD_{23} \cD_{12} \cos\PhiA_{L2} }
            { \cD_{23} \cD_{13} \cos\PhiA_{L3}
             -\sD_{23} \sD_{12} \sD_{13} \cos(\PhiA_{L3} - \CPVD) }
\label{eq:Append_D_DA_44}
\end{eqnarray}
\begin{eqnarray}
& & \hskip-1.2cm
  \cos \thA_{12} = 
       \frac{ \cD_{23} \sD_{12} \cD_{13} \cos(\PhiA_{15} - \CPVD)
             -\sD_{23} \sD_{13} \cos\PhiA_{15} }  
            { \cD_{12} \cD_{13} \cos(\PhiA_{L1} + \PhiA_{R1}) } 
\label{eq:Append_D_DA_45}
\end{eqnarray}
\begin{eqnarray}
& & \hskip-1.2cm
  \cos \thA_{13} = 
      \frac{ \cD_{12} \cD_{13} \cos(\PhiA_{L1} + \PhiA_{R1}) }
           { \cA_{12} }
\label{eq:Append_D_DA_46}
\end{eqnarray}
The remaining parameters thus can be determined as follows.
\begin{eqnarray}
& & \hskip-1.2cm
  \sin^2{\PhiA_{R1}}     \nonumber \\
& & \hskip-1.2cm
      = \frac{  (\sin\PhiA_{L3} - d \sin\CPVD)^2 }
            {  (\sin\PhiA_{L3} - d \sin\CPVD)^2  
             + (d \cos\CPVD - \cos\PhiA_{L3})^2 }  ,
\label{eq:Append_D_DA_47}
\end{eqnarray}
where 
\begin{equation*}
d = \frac{ \cA_{23} \sD_{12} }{ \sA_{23} \cD_{12} \sD_{13} }  .
\end{equation*}
Therefore, 
$\PhiA_{L1} = -\PhiA_{R1}$ and  
$\PhiA_{R2} = \PhiA_{15} - \PhiA_{L1}$ can be determined.
Finally, the CP-violating phase in representation $\RepA$, $\CPVA$, can be 
resolved using those conditions associated with $\sin \CPVA$:
\begin{eqnarray}
& & \hskip-1.0cm 
  \sin \CPVA                                \nonumber \\
& & \hskip-1.0cm
  = \frac{ \sD_{12} \sin(\PhiA_{L2} + \PhiA_{R1} + \CPVD) }
         { \sA_{23} \cA_{12} \sA_{13} }     \nonumber \\   
& & \hskip-1.0cm
  = \frac{ \cD_{12} \sD_{13} \sin(\PhiA_{L3} + \PhiA_{R1}) }
         { \cA_{23} \cA_{12} \sA_{13} }     \nonumber \\   
& & \hskip-1.0cm
  = \frac{-\cD_{23} \cD_{12} \sin(\PhiA_{L2} + \PhiA_{R2}) }
         { \sA_{23} \sA_{12} \sA_{13} }     \nonumber \\   
& & \hskip-1.0cm
  = \frac{ \cD_{23} \sD_{12} \sD_{13} \sin(\eta - \CPVD)
          +\sD_{23} \cD_{13} \sin\eta } 
         { \cA_{23} \sA_{12} \sA_{13} }     \nonumber \\   
& & \hskip-1.0cm
  = \frac{ \sD_{23} \sD_{12} \cD_{13} \sin(\PhiA_{L1} - \CPVD)
          +\cD_{23} \sD_{13} \sin\PhiA_{L1} }
         { \sA_{13} }    
\label{eq:Append_D_DA_5}
\end{eqnarray}
where $\eta \equiv \PhiA_{L3} + \PhiA_{R2}$. 
Consistency in the value of $\sin \CPVA$ calculated from the five different 
expressions in Eq.(\ref{eq:Append_D_DA_5}) serves as a means to check whether 
the values of the nine parameters are correct. 


%



\begin{thebibliography}{99}

\bibitem{Nakamura:2010zzi}
  K.~Nakamura {\it et al.} [ Particle Data Group Collaboration ],
  J.\ Phys.\ G {\bf G37}, 075021 (2010).

\bibitem{Maki:1962mu}  
  Z.~Maki, M.~Nakagawa and S.~Sakata,
  Prog.\ Theor.\ Phys.\  {\bf 28}, 870 (1962).

\bibitem{Pontecorvo:1967fh}  
  B.~Pontecorvo,
  Sov.\ Phys.\ JETP {\bf 26}, 984 (1968)
  [Zh.\ Eksp.\ Teor.\ Fiz.\  {\bf 53}, 1717 (1967)].

\bibitem{Schechter:1980gr}  
  J.~Schechter and J.~W.~F.~Valle,
  Phys.\ Rev.\  D {\bf 22}, 2227 (1980).


\bibitem{Fritzsch:1997st}
  H.~Fritzsch, Z.~-z.~Xing,
  Phys.\ Rev.\  {\bf D57}, 594-597 (1998) [hep-ph/9708366].

\bibitem{Choubey:2000bf}
  S.~Choubey, S.~Goswami, K.~Kar,
  Astropart.\ Phys.\  {\bf 17}, 51-73 (2002) [arXiv:hep-ph/0004100 [hep-ph]].

\bibitem{Giunti_Kim}
    Carlo Giunti and Chung W. Kim, \emph{Fundamentals of Neutrino Physics
    and Astrophysics}, published by Oxford University Press Inc. New York
    (2007).

\bibitem{Zheng:2010kp}  
  Y.~j.~Zheng,
  Phys.\ Rev.\  D {\bf 81}, 073009 (2010) [arXiv:1002.0919 [hep-ph]].

\bibitem{Huang:2011by}
  M.~Huang, D.~Liu, J.~C.~Peng, S.~D.~Reitzner and W.~C.~Tsai,
  arXiv:1108.3906 [hep-ph].

\bibitem{Aguilar:2001ty}
  A.~Aguilar {\it et al.}  [LSND Collaboration],
  Phys.\ Rev.\  D {\bf 64}, 112007 (2001) [arXiv:hep-ex/0104049].

\bibitem{Chau:1984fp}
  L.~-L.~Chau, W.~-Y.~Keung,
  Phys.\ Rev.\ Lett.\  {\bf 53}, 1802 (1984).

\bibitem{Aharmim:2009gd} 
  B.~Aharmim {\it et al.}  [SNO Collaboration],
  Phys.\ Rev.\  C {\bf 81}, 055504 (2010) [arXiv:0910.2984 [nucl-ex]].

\bibitem{Aharmim:2011vm}  
   {\it et al.}  [SNO Collaboration],
  [arXiv:1109.0763 [nucl-ex]].

\bibitem{Hosaka:2005um}  
  J.~Hosaka {\it et al.}  [Super-Kamkiokande Collaboration],
  Phys.\ Rev.\  D {\bf 73}, 112001 (2006) [arXiv:hep-ex/0508053].

\bibitem{Cravens:2008zn}  
  J.~P.~Cravens {\it et al.}  [Super-Kamiokande Collaboration],
  Phys.\ Rev.\  D {\bf 78}, 032002 (2008) [arXiv:0803.4312 [hep-ex]].

\bibitem{Gando:2010aa}  
  A.~Gando {\it et al.}  [The KamLAND Collaboration],
  Phys.\ Rev.\  D {\bf 83}, 052002 (2011) [arXiv:1009.4771 [hep-ex]].

\bibitem{Apollonio:2002gd}  
  M.~Apollonio {\it et al.}  [CHOOZ Collaboration],
  Eur.\ Phys.\ J.\  C {\bf 27}, 331 (2003) [arXiv:hep-ex/0301017].

\bibitem{Ahn:2006zza}  
  M.~H.~Ahn {\it et al.}  [K2K Collaboration],
  Phys.\ Rev.\  D {\bf 74}, 072003 (2006) [arXiv:hep-ex/0606032].

\bibitem{Adamson:2011ig}  
  P.~Adamson {\it et al.}  [The MINOS Collaboration],
  Phys.\ Rev.\ Lett.\  {\bf 106}, 181801 (2011) [arXiv:1103.0340 [hep-ex]].

\bibitem{MINOS_nue_app_2011}    
   L. Whitehead [for MINOS Collaboration], “Recent results from MINOS”,
      Joint Experimental-Theoretical Seminar (24 June 2011, Fermilab, USA);
      websites: theory.fnal.gov/jetp, http://www-numi.fnal.gov/pr\_plots/

\bibitem{Wendell:2010md}
  R.~Wendell {\it et al.} [Kamiokande Collaboration],
  Phys.\ Rev.\  {\bf D81}, 092004 (2010),  [arXiv:1002.3471 [hep-ex]].

\bibitem{Aharmim:2009zm}
  B.~Aharmim {\it et al.}  [SNO Collaboration],
  Phys.\ Rev.\  D {\bf 80}, 012001 (2009), [arXiv:0902.2776 [hep-ex]].

\bibitem{Hosaka:2006zd}
  J.~Hosaka {\it et al.} [Super-Kamiokande Collaboration],
  Phys.\ Rev.\  {\bf D74 }, 032002 (2006),  [hep-ex/0604011].

\bibitem{Fogli:2011qn}
  G.~L.~Fogli, E.~Lisi, A.~Marrone, A.~Palazzo, A.~M.~Rotunno,
     [arXiv:1106.6028 [hep-ph]].

\bibitem{Schwetz:2011qt}
  T.~Schwetz, M.~Tortola, J.~W.~F.~Valle,
  New J.\ Phys.\  {\bf 13}, 063004 (2011),  [arXiv:1103.0734 [hep-ph]].

\bibitem{GonzalezGarcia:2010er}
  M.~C.~Gonzalez-Garcia, M.~Maltoni, J.~Salvado,
  JHEP {\bf 1004}, 056 (2010),  [arXiv:1001.4524 [hep-ph]].

\bibitem{Roa:2009wp}
  J.~E.~Roa, D.~C.~Latimer, D.~J.~Ernst,
  Phys.\ Rev.\  {\bf C81}, 015501 (2010), [arXiv:0904.3930 [nucl-th]].

\bibitem{Balantekin:2008zm}
  A.~B.~Balantekin, D.~Yilmaz,
     J.\ Phys.\ G {\bf G35}, 075007 (2008), [arXiv:0804.3345 [hep-ph]].

\bibitem{Ge:2008sj}
  H.~L.~Ge, C.~Giunti, Q.~Y.~Liu,
  Phys.\ Rev.\  {\bf D80}, 053009 (2009),  [arXiv:0810.5443 [hep-ph]].

\bibitem{Goswami:2004cn}
  S.~Goswami and A.~Y.~Smirnov,
  Phys.\ Rev.\  D {\bf 72}, 053011 (2005), [arXiv:hep-ph/0411359].

\bibitem{Choubey:2003uw}
  S.~Choubey,
  J.\ Phys.\ G {\bf G29}, 1833-1837 (2003).

\bibitem{Guo:2007ug}
  X.~Guo {\it et al.} [Daya Bay Collaboration],
  [hep-ex/0701029].

\bibitem{Ardellier:2006mn}
  F.~Ardellier {\it et al.} [Double Chooz Collaboration],
  [hep-ex/0606025].

\bibitem{Joo:2007zzb}
  K.~K.~Joo [ RENO Collaboration ],
  Nucl.\ Phys.\ Proc.\ Suppl.\  {\bf 168}, 125-127 (2007).

\bibitem{Anjos:2005pg}
  J.~C.~Anjos {\it et al.} [Angra Collaboration],
  Nucl.\ Phys.\ Proc.\ Suppl.\  {\bf 155}, 231 (2006),
  [arXiv:hep-ex/0511059].

\bibitem{Chen:2008eq}
  M.~-C.~Chen, K.~T.~Mahanthappa,
  Nucl.\ Phys.\ Proc.\ Suppl.\  {\bf 188}, 315-320 (2009).
  [arXiv:0812.4981 [hep-ph]].

\bibitem{double_chooz_2011}  
  http://doublechooz.in2p3.fr/Status\_and\_News/status\_and\_news.php

\bibitem{T2K_expt}
  http://j-parc.jp/NuclPart/pac\_0606/pdf/p11-Nishikawa.pdf

\bibitem{Ayres:2004js}
  D.~S.~Ayres {\it et al.}  [NOvA Collaboration],
  [arXiv:hep-ex/0503053].

\bibitem{Itow:2001ee}
  Y.~Itow {\it et al.}  [The T2K Collaboration],
  [arXiv:hep-ex/0106019].

\bibitem{Hagiwara:2006vn}
  K.~Hagiwara, N.~Okamura and K.~i.~Senda,
  Phys.\ Rev.\  D {\bf 76}, 093002 (2007), [arXiv:hep-ph/0607255].

\bibitem{Jarlskog:1985ht}
  C.~Jarlskog,
  Phys.\ Rev.\ Lett.\  {\bf 55}, 1039 (1985).

\bibitem{Wu:1985ea}
  D.~-d.~Wu,
  Phys.\ Rev.\  {\bf D33}, 860 (1986).

\bibitem{Wolfenstein_msw}
    L. Wolfenstein, Phys. Rev. D \textbf{17}, 2369 (1978).

\bibitem{Mik_Smir_msw}
    S. P. Mikheyev and A.Yu. Smirnov,
       Sov. J. Nucl. Phys. \textbf{42}, 913 (1985).

\bibitem{Cleveland:1998nv}  
  B.~T.~Cleveland {\it et al.},
  Astrophys.\ J.\  {\bf 496}, 505 (1998).

\bibitem{Abdurashitov:2009tn} 
  J.~N.~Abdurashitov {\it et al.}  [SAGE Collaboration],
  Phys.\ Rev.\  C {\bf 80}, 015807 (2009), [arXiv:0901.2200 [nucl-ex]].

\bibitem{Arpesella:2008mt}  
  C.~Arpesella {\it et al.}  [The Borexino Collaboration],
  Phys.\ Rev.\ Lett.\ {\bf 101}, 091302 (2008), [arXiv:0805.3843 [astro-ph]].

\bibitem{Mueller:2011nm}  
  T.~.A.~Mueller, D.~Lhuillier, M.~Fallot, A.~Letourneau, S.~Cormon, 
  M.~Fechner, L.~Giot, T.~Lasserre {\it et al.},
  Phys.\ Rev.\  {\bf C83}, 054615 (2011),  [arXiv:1101.2663 [hep-ex]].

\bibitem{Solar_Nu_Eng_Bahcall}  
    http://www.sns.ias.edu/~jnb/SNdata/sndata.html

\bibitem{Winter:2004kf}  
  W.~T.~Winter, S.~J.~Freedman, K.~E.~Rehm and J.~P.~Schiffer,
  Phys.\ Rev.\  C {\bf 73}, 025503 (2006), [arXiv:nucl-ex/0406019].

\bibitem{Bahcall:2004pz}  
  J.~N.~Bahcall, A.~M.~Serenelli and S.~Basu,
  Astrophys.\ J.\  {\bf 621}, L85 (2005), [arXiv:astro-ph/0412440].

\bibitem{Serenelli:2009yc}  
  A.~Serenelli, S.~Basu, J.~W.~Ferguson and M.~Asplund,
  Astrophys.\ J.\  {\bf 705}, L123 (2009), [arXiv:0909.2668 [astro-ph.SR]].  

\bibitem{PEM-C}
    A. Dziewonski, 
       Physics of the Earth and Planetary Interiors {\bf 10}, 12 (1975).
\bibitem{PREM}
    A. M. Dziewonski and D. L. Anderson, 
       Physics of the Earth and Planetary Interiors {\bf 25}, 297 (1981).
\bibitem{Ioannisian_2004}
   A. N. Ioannisiana and A. Yu. Smirnov, 
      Phys. Rev. Lett. {\bf 93}, 241801 (2004).

\bibitem{Casper:2002sd}
  D.~Casper,
  Nucl.\ Phys.\ Proc.\ Suppl.\  {\bf 112}, 161 (2002), [arXiv:hep-ph/0208030].
\bibitem{Barger:1980tf}
  V.~D.~Barger, K.~Whisnant, S.~Pakvasa and R.~J.~N.~Phillips,
  Phys.\ Rev.\  D {\bf 22}, 2718 (1980).

\bibitem{Aharmim:2005gt}
  B.~Aharmim {\it et al.}  [SNO Collaboration],
  Phys.\ Rev.\  C {\bf 72}, 055502 (2005), [arXiv:nucl-ex/0502021].

\bibitem{Fogli:1994nn} 
  G.~L.~Fogli and E.~Lisi,
  Astropart.\ Phys.\  {\bf 3}, 185 (1995).

\bibitem{Fogli:1999zg}  
  G.~L.~Fogli, E.~Lisi, D.~Montanino and A.~Palazzo,
  Phys.\ Rev.\  D {\bf 62}, 013002 (2000), [arXiv:hep-ph/9912231].

\bibitem{Garzelli:2000tn}  
  M.~V.~Garzelli and C.~Giunti,
  Phys.\ Lett.\  B {\bf 488}, 339 (2000), [arXiv:hep-ph/0006026].

\bibitem{Garzelli:2001zu}  
  M.~V.~Garzelli and C.~Giunti,
  JHEP {\bf 0112}, 017 (2001), [arXiv:hep-ph/0108191].

\bibitem{Fogli:2005qa}  
  G.~L.~Fogli, E.~Lisi, A.~Palazzo and A.~M.~Rotunno,
  Phys.\ Lett.\  B {\bf 623}, 80 (2005), [arXiv:hep-ph/0505081].

\bibitem{KL_2nd_fission_flux}   
   http://www.awa.tohoku.ac.jp/KamLAND/datarelease/fission\_flux\_distance.dat

\bibitem{Vogel:1999zy}  
  P.~Vogel and J.~F.~Beacom,
  Phys.\ Rev.\  D {\bf 60}, 053003 (1999), [arXiv:hep-ph/9903554].

\bibitem{Fogli:2003th}  
  G.~L.~Fogli, E.~Lisi, A.~Marrone and D.~Montanino,
  Phys.\ Rev.\  D {\bf 67}, 093006 (2003), [arXiv:hep-ph/0303064].

\bibitem{Wendell:2008zz} 
  R.~A.~Wendell, Ph.D. Thesis, University of North Carolina (2008).

\bibitem{Ashie:2005ik}
  Y.~Ashie {\it et al.}  [Super-Kamiokande Collaboration],
  Phys.\ Rev.\  D {\bf 71} (2005) 112005 [arXiv:hep-ex/0501064].

\bibitem{Honda:2004yz}
  M.~Honda, T.~Kajita, K.~Kasahara and S.~Midorikawa,
  Phys.\ Rev.\  D {\bf 70} (2004) 043008 [arXiv:astro-ph/0404457].

\bibitem{Fogli:2002pt}
  G.~L.~Fogli, E.~Lisi, A.~Marrone, D.~Montanino and A.~Palazzo,
  Phys.\ Rev.\  D {\bf 66} (2002) 053010 [arXiv:hep-ph/0206162].

\bibitem{Sonley:2009}
  T.~J.~Sonley,
  PhD thesis (2009)

\bibitem{Barr:2004br}
  G.~D.~Barr, T.~K.~Gaisser, P.~Lipari, S.~Robbins and T.~Stanev,
  Phys.\ Rev.\  D {\bf 70} (2004) 023006 [arXiv:astro-ph/0403630].
\end{thebibliography}
\end{document}